\documentclass[prb,amsmath,superscriptaddress,twocolumn,showpacs]{revtex4-1}
\usepackage{bm}
\usepackage{amssymb}
\usepackage{colordvi}
\usepackage{graphicx}
\usepackage{color}
\usepackage{hyperref}

\newcommand{\ov}[1]{\overline{{#1}}}
\newcommand{\be}{\begin{equation}}
\newcommand{\ee}{\end{equation}}
\newcommand{\bea}{\begin{eqnarray}}
\newcommand{\eea}{\end{eqnarray}}

\newcommand{\vep}{\varepsilon}
\newcommand{\ave}[1]{\left\langle #1\right\rangle}
\newcommand{\ome}{\omega}

\def\nn{\nonumber}

\begin{document}

\title{Hopping thermoelectric transport in finite systems: boundary
  effects}

\author{Jian-Hua Jiang}
\affiliation{Department of Condensed Matter Physics, Weizmann Institute of
  Science, Rehovot 76100, Israel}
\affiliation{Department of Physics, University of Science and Technology of China, Hefei, Anhui, 230026, China}
\author{Ora Entin-Wohlman}
  \affiliation{Raymond and Beverly Sackler School of Physics and Astronomy, Tel Aviv University, Tel Aviv 69978, Israel}\affiliation{Department of Physics and the Ilse Katz Center for Meso-
  and Nano-Scale Science and Technology, Ben Gurion University, Beer
  Sheva 84105, Israel}
\author{Yoseph Imry}
\affiliation{Department of Condensed Matter Physics, Weizmann Institute of
  Science, Rehovot 76100, Israel}

\date{\today}

\begin{abstract}
It is shown that for the hopping regime, the thermopowers in both finite two-terminal and three-terminal systems are governed by the edges of the samples. This is due to the fact that the energy transfer between a transport electron and a conducting terminal
is determined by the  site most strongly coupled to that terminal. One-dimensional systems with both nearest-neighbor and variable-range transport as well as certain types of two-dimensional systems, are considered. For a given sample, the changes in the thermopowers due to modifying the bulk are quite  limited, compared with those of the conductance. When the small thermopower changes exist, their average over a large ensemble  of mesoscopic samples will vanish. We also obtain the distribution of the thermopower in  such an  ensemble and show that its width approaches a finite limit with increasing sample length. This contrasts with the distribution of conductances in such systems, whose width vanishes in the long sample limit. Finally, we find that the thermal conductances in the three-terminal case have a boundary-dominated contribution, due to non-percolating conduction paths. This contribution can become dominant when the usual conductance is small enough. All our theoretical statements are backed by numerical computations.

\end{abstract}

\pacs{ 72.20.Ee, 72.20.Pa, 84.60.Rb}

\maketitle

\section{Introduction}


Achieving   large thermopowers is a challenge to our understanding of  electronic transport. 
At the same time, it  is a crucial ingredient \cite{Gen,ali} for many energy-conversion devices.
In good, wide band, conductors the thermopower, S, is typically very small, due to the approximate 
electron-hole symmetry. Breaking this symmetry is therefore important for obtaining large values of S.
This happens in various poor  or narrow band \cite{mahan} conductors/semiconductors, 
near the metal-insulator transition \cite{mottcutler,Joesivan,ariel,mit,2dmit} and in the hopping regime. 
\cite{ZO,zvyaginbook}

Recent experiments addressed
thermoelectric transport in mesoscopic systems.\cite{nanoexp,meso-rmp}   Besides
their general interest, they may be relevant for small-scale thermoelectric devices. 
\cite{frige,meso-rmp,ali} Especially in the hopping regime, where
the electronic states are localized and discrete, electron-hole symmetry is usually 
broken in a given sample, even if it is restored by averaging over many realizations. This should result in relatively large, sample-specific, thermopower.
In addition, 
the parasitic phonon heat conductivity can be reduced due to interfaces and sample shape and geometry effects. 
\cite{low-d,interfaces}

\begin{figure}[htb]
  \centering \includegraphics[height=3.8cm]{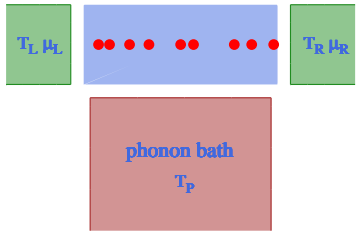}
  \caption{(Color online) Schematic of the three-terminal geometry. 
    A 1D system (blue region) consisting of a number of LSs is
    connected to three terminals: the left and right ones (green)
    are electronic leads which have their own temperatures and
    chemical potentials, the terminal below (brown) is a thermal
    terminal (a phonon bath in this work) at temperature
    $T_P$. }
\label{fig1}
\end{figure}

Most of the
studies on the thermoelectric effects in the hopping regime were devoted to
bulk systems.\cite{mottcutler,ZO,zvyaginbook}   Recently, we
discussed the thermoelectric transport properties in finite
one-dimensional (1D) systems\cite{ourprb} where the boundary effect was
found to be very important for the thermopower.
In this work we follow up, substantiate and generalize that study. The importance of edge effects on the thermopowers will be highlighted. We shall consider  
both 1D and 2D finite systems, which can be arbitrarily large. Since  the conduction electrons have to exchange 
energy with a reservoir, the ``three-terminal
geometry''\cite{3t,3t1,ourprb} naturally appears.  In addition to the two electronic 
terminals which exchange charge and energy/heat, the third terminal is 
purely thermal and mainly exchanges energy with the conduction electrons.
The three-terminal setup for 1D finite
systems is shown in Fig.~\ref{fig1}. The system, bridging two electronic terminals 
(leads) consists of a
number of localized states (LSs) with random energies. The system is
connected to the leads by (dominantly elastic) tunnel couplings. 
Electronic conduction through the system is achieved via tunneling and
phonon-assisted hopping. The setup can be realized, e.g., when the two 
 electronic leads are suspended and the system is mounted
 on a (boson bath) substrate. \cite{Com}  The complete description of the
thermoelectric transport in the linear-response regime is a $3\times
3$ transport matrix relating the three currents to three
``forces'' (or ``affinities"),\cite{3t,3t1,ourprb} see the next section.

In this paper we consider a noninteracting localized system
with constant density of states and localization length, in an energy
window $(-E_{c},E_{c})$.\cite{note1}
We start with  a short review of the three-terminal thermoelectric transport 
(Sec. \ref{ONS}).  We 
show that the boundary effect
dominates the thermopower in nearest-neighbor hopping (NNH) 1D systems (Sec. \ref{BOUND}). 
The 
simple underlying physics is illustrated via the solution of a 
simple 3-site model (Sec. \ref{BOUNDA} and Appendix \ref{ap:prob}). Then, longer 
NNH1D systems 
are considered (Sec. \ref{LONGER}). In Sec. \ref{SIMPLIFIED},
simplified types of 2D systems are treated. Because of incomplete
averaging at the boundaries there can be a {\em finite and fluctuating}
thermopower even for a very large system.  The situation
for variable-range hopping (VRH) is discussed in Sec. \ref{VRHSEC}. 
In Sec.  \ref{VRHS} we
shall first focus  on the 1D case and then consider small   2D systems, whose  width and length are comparable.
In these cases the bulk effect plays a limited role, which will  tend to disappear when the
system size increases.  When the system size is increased  further, one would expect 
both the thermopower and its fluctuations to  eventually go to zero in the macroscopic
limit as a consequence of the  particle-hole symmetry being restored with 
averaging. We find however that the fluctuations remain finite for an arbitrarily long system, as long as its transverse size  is finite.
Finally, we discuss special effects of  ``non-spanning electronic paths" (that do not 
transport charge) to the thermal conductances (Sec. \ref{SPECIAL}). Our statements 
are backed up numerically. The numerical scheme is explained 
in Appendix \ref{app:num}, and the contributions of various conducting paths  are compared in 
Appendix \ref{ap:tun}.
Finally, Sec. \ref{CD} includes a short summary and conclusions.

\section{Thermoelectric Transport through Localized states in
  the three-terminal geometry}
  
  \label{ONS}

The study of hopping thermoelectric transport in three-terminal
geometry was done in Refs.~\onlinecite{ourprb,3t}. For completeness we
summarize here the basic formulation of the problem. The hopping transitions
between the LSs are assisted by phonons. We focus on the situation
when the on-site Coulomb interaction is so strong that each LS can
only be occupied by at most one electron. The inter-site Coulomb
interaction may lead to interesting effects but these will not be discussed
here.\cite{note3} The hopping rate from LS $i$ to LS $j$ for, say,
$\vep_j>\vep_i$, is given by the Fermi golden-rule 
as 
\bea
\Gamma_{i\to j} &=& 2\pi \sum_{\bf q}|\alpha_{ij{\bf
  q}}|^2\delta(\vep_j-\vep_i-\ome_{\bf q})f_i(1-f_j) N_{ij},\nn\\
&\equiv& \gamma_{ij} f_i(1-f_j)N_{ij}\  .\label{g1}
\eea
Here, $\alpha_{ij{\bf q}}$ is the electron-phonon interaction matrix
element between the two LSs, $\vep_j$ and $\vep_i$ are the energy of
the LSs $j$ and $i$, respectively, $\ome_{\bf q}$ is the phonon energy,
$f_j$ and $f_i$ are the occupation probability on the LSs $j$ and $i$,
respectively,  and $N_{ij}$ is the phonon distribution at the energy
$\ome_{\bf q}=|\vep_j-\vep_i|$. The phonon distribution in the system
is determined by the phonon bath,  $N_{ij} =[\exp (|\varepsilon_j-\varepsilon_i |/(k_{B}T_{P}))-1]^{-1}$.
At large distances the overlap of the wavefunctions of the two LSs is
exponentially small. The asymptotic behavior of $\gamma_{ij}$ is thus
$\gamma_{ij} \sim \gamma_{ep} \exp(-2r_{ij}/\xi)$ where
$r_{ij}=|{\bf r}_j-{\bf r}_i|$ is the distance between the LSs with
 ${\bf r}_j$ and ${\bf r}_i$ denoting their position vectors, $\xi$ 
is the localization length, and $\gamma_{ep}$ is proportional to the
electron-phonon coupling and the  phonon density of states.
The hopping rate from LS $i$ to the left lead is 
\be
\Gamma_{i\to L} = \gamma_{iL} f_i(1-f_L(\vep_i))\ ,  \label{g2}
\ee
where $\gamma_{iL}=2\pi|\alpha_{iL}|^2\rho_L$ with
$\alpha_{iL}$ standing for the coupling between the LS $i$ and the extended
states in the left lead of which $\rho_L$  is the density of
states. We focus on the situation where $\gamma_{iL}$ does not depend
on the energy $\vep_i$ (i.e., no particle-hole asymmetry). $f_L$
stands for the distribution in the left lead, $f_{L}(\varepsilon )=[\exp (\varepsilon -\mu_{L})/(k_{B}T_{L}))+1]^{-1}$.
The transition rate from 
$i$ to the right lead can be written down similarly. The asymptotic 
behavior of $\gamma_{iL}$ and $\gamma_{iR}$ at large distances is also
exponential, $\gamma_{iL}\sim \gamma_e\exp(-2r_{iL}/\xi)$
[$\gamma_{iR}\sim \gamma_e \exp(-2r_{iR}/\xi)$] where $r_{iL}$
[$r_{iR}$] is the distance between LS $i$ and the left (right) lead
and $\gamma_e$ scales with the tunnel coupling strength between LSs
and the leads. The electric current flowing from $i$ to $j$ is
\be
I_{i\to j} = e(\Gamma_{i\to j} - \Gamma_{j\to i})\ , \label{g3}
\ee
with $e$ being the charge  of the carrier. 
The electric current flowing from $i$ to the left (right) lead is
calculated similarly. At steady state, according to  Kirchhoff's
current law,
\be
\sum_{j} I_{i\to j} + I_{i\to L} + I_{i\to R}= 0 . \label{g4}
\ee
which is also a statement of probability conservation. The steady-state distribution on each LS is obtained by solving Eqs.~(\ref{g1}),
(\ref{g2}), (\ref{g3}), and (\ref{g4}).

Before formulating the currents among the three terminals in terms of the
transition rates, we  present a thermodynamic
analysis in the linear-response regime. There are
three heat currents flowing into  each terminal, $\dot Q_L$,
$\dot Q_R$, and $\dot Q_P$, as well as two particle currents flowing
into the electronic terminals, $\dot N_L$ and $\dot N_R$. The heat
currents are related to the energy and particle currents according to
the thermodynamic relation $\dot Q_i = \dot E_i - \mu_i\dot N_i$ for
$i=L,R$ and $\dot Q_P=\dot E_P$ for the phononic terminal with $\dot
E_i$ being the energy current flowing into terminal $i$. Particle and
energy conservation renders $\sum_{i=L,R} \dot N_i =0$,
$\sum_{i=L,R,P}\dot E_i =0$. Hence there are only three independent
currents which are the electric current $I_e=e\dot N_R=-e\dot N_L$ and
two heat currents. For the latter one can choose $\dot Q_L$ and $\dot
Q_R$, or any two linear independent combinations of them. We shall
adopt the convention introduced  in our previous work\cite{ourprb} and choose the
following heat currents \cite{ZO,zvyaginbook}
\be
I_Q^e = \frac{1}{2}(\dot Q_R - \dot Q_L), \quad I_{Q}^{pe} = -\dot Q_P =
\dot Q_L + \dot Q_R\  .\label{para}
\ee
In the linear-response regime the entropy production rate
$\dot S$ is given by
$T\dot{S}=T[(\dot{Q}_{L}/T_{L})+(\dot{Q}_{R}/T_{R})+\dot{Q}_{P}/T_{P})]=I_{e}(\delta\mu/e)+I^{e}_{Q}(\delta /T)+I^{pe}_{Q}(\Delta T/T)$, where $T$ is the common (equilibrium) temperature of the setup. This relation
identifies the three ``forces'' (affinities) conjugated to the three currents, 
$\delta\mu=\mu_L-\mu_R$,   $\delta T=T_L-T_R$, and
 $\Delta T = T_P-(T_L+T_R)/2$.
The phenomenological linear-transport equation which satisfies the
Onsager reciprocity relations is then\cite{ourprb} 
\begin{align}
\left( \begin{array}{c}
I_e\\ I_Q^e\\ I_{Q}^{pe}\end{array}\right) =
\left( \begin{array}{cccc}
G & L_1 & L_2 \\
L_1 & K_e^0 & L_3 \\
L_2 & L_3 & K_{pe} 
\end{array}\right) \left(\begin{array}{c}
\delta\mu/e\\ \delta T/T\\ \Delta T/T \end{array}\right)\  ,
\label{trans}
\end{align}
In the three-terminal geometry, besides the normal thermopower $S=L_{1}/(TG)$
there is the three-terminal thermopower $S_p = L_2/(TG)$ which
converts the temperature difference $\Delta T$ to voltage (and vice
versa).\cite{ourprb,3t1} 

We now formulate the currents $I_e$, $I_Q^e$, and $I_Q^{pe}$ in terms
of microscopic quantities. The electric current is given by
\bea
I_e = -\sum_i I_{i\to L} = \sum_i I_{i\to R}\  .\label{is1}
\eea
The heat currents $I_Q^e$ and $I_Q^{pe}$ can be obtained from $\dot
Q_L$ and $\dot Q_R$, 
\bea
\hspace{-0.3cm} \dot Q_L = \sum_i \left(\frac{E_i-\mu}{e}\right) I_{i\to L}, \
 \dot Q_R = \sum_i\left(\frac{E_i-\mu}{e}\right) I_{i\to R}  \label{is2}
\eea
where $\mu$ is 
the equilibrium chemical potential. 

In the linear-response regime the current between two LSs can be
written as
\be
I_{i\to j} = G_{ij}( U_i - U_j \pm U_{ij} ) \ ,\label{gg1}
\ee
where the signs $+$ and $-$ are for $\vep_j>\vep_i$ and
$\vep_j<\vep_i$,  respectively, and  the conductance of the bond $(ij)$ is $G_{ij}=(e^{2}/(k_{B}T))\gamma_{ij}
f_i^0(1-f_j^0) (N_{ij}^0+1/2\mp 1/2)$
 (the superscript 0 is used to denote an equilibrium distribution function). In Eq. (\ref{gg1})
$U_{i}=(k_{B}T/e)(f_{i}-f^{0}_{i})/(f^{0}_{i}(1-f^{0}_{i}))$
and $U_{ij}=(k_{B}T/e)(N_{ij}-N^{0}_{ij})/(N^{0}_{ij}(N^{0}_{ij}+1))$.
To complete the description of the  resistor
network we also write down the current between the leads and the LSs 
\be
I_{i\to L} = e(\Gamma_{i\to L}-\Gamma_{L\to i}) = G_{iL}\left[
  U_i - U_L(\vep_i) \right]\ , \label{gg2} 
\ee
where $G_{iL} =(e^{2}/(k_{B}T)) 
\gamma_{iL}f_i^0(1-f_i^0)$ and          
$U_{L}(\varepsilon_{i})=(k_{B}T/e)[f_{L}(\epsilon_{i})-f^{0}_{i}]/(f^{0}_{i}(1-f^{0}_{i}))$.
We shall adopt the widely-used approximation, valid in the rather broad regimes discussed in Refs. \onlinecite{MA,AHL,arielvrh}, which yields
\bea
G_{ij} &\simeq& G_0
\exp\left(-\frac{2r_{ij}}{\xi}-\frac{|\vep_i-\mu|+|\vep_j-\mu|+|\vep_i-\vep_j|}{2k_BT}\right),\nn\\
G_{iL} &\simeq& G_0\exp\left(-\frac{2r_{iL}}{\xi}-\frac{|\vep_i-\mu|}{k_BT}\right),\label{gijl}
\eea
where
$G_0\sim e^2\gamma_{ep}/(k_BT)\sim
e^2\gamma_{ei}/(k_BT)$ has been introduced.
This coefficient sets the scale of the whole conductance and will not 
play a role in the subsequent
discussions. 
The resistor network described above has been investigated for the case  $\Delta T=0$ a long time ago. \cite{note2}
We have recently considered the  effect of the
term $U_{ij}$, which arises due to a finite small $\Delta T$. \cite{
ourprb}. A scheme for the numerical solution of the
resistor network is presented in Appendix~\ref{app:num}. The above
formalism reduces to the Miller-Abrahams resistor network
model\cite{MA,AHL} when there is no temperature difference.

\section{Boundary effect in 1D NNH systems}
\label{BOUND}

\subsection{A simple three-site 1D NNH system}

\label{BOUNDA}

\begin{figure}[htb]
  \centering\hspace{0.5cm} \includegraphics[height=3.5cm]{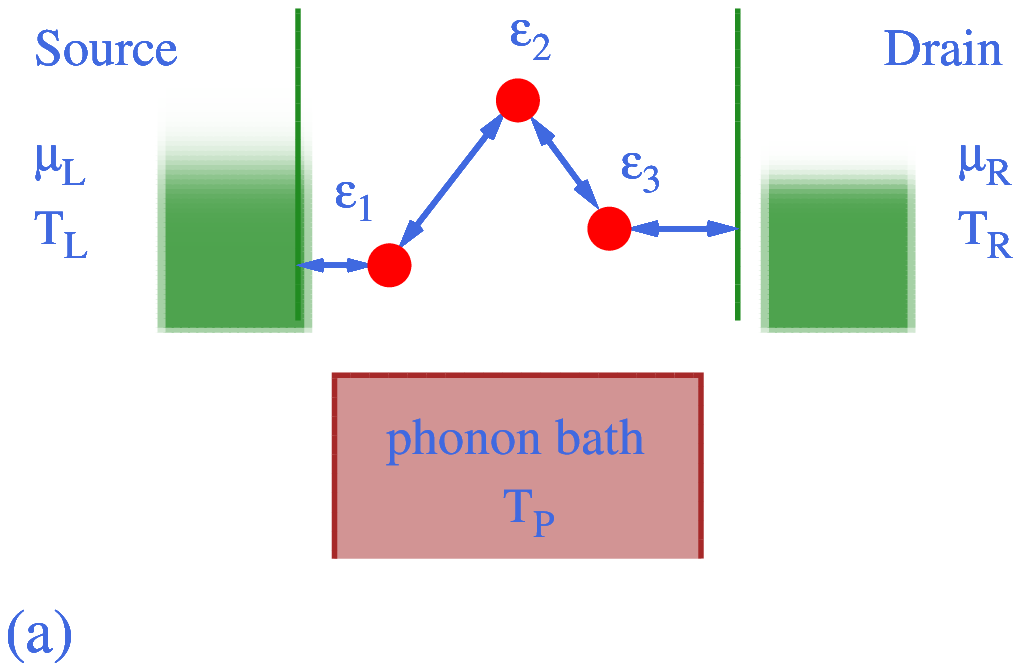}\\
  \centering \includegraphics[height=3.8cm]{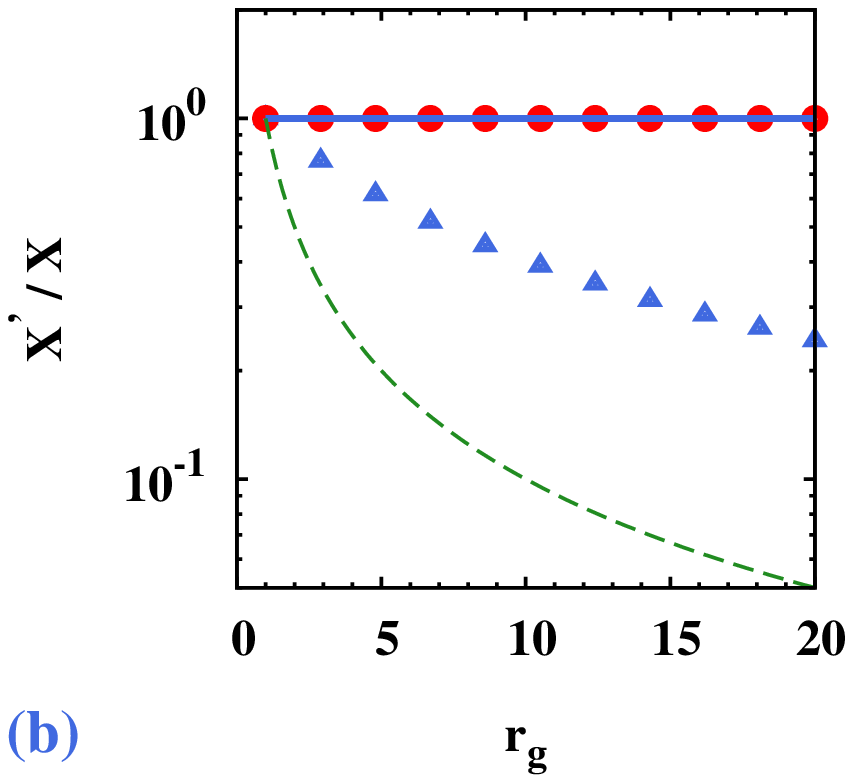}
  \caption{(Color online) 1D NNH. (a): Schematic of the three-site model in the
    three-terminal geometry. The three red dots stand for the three
    LSs. The arrows represent the hopping transitions in the dominant
    hopping path. The vertical (horizontal) direction
    stands for the energy (position). The color densities in the left
    and right sides represent the electronic distribution in the left
    and right leads,  respectively. In the figure the left lead has
    higher chemical potential and temperature. (b) The effect of
   decreasing the conductances in the middle of the system by a
    factor  $r_g$ on the conductance and thermopower of a 1D NNH system. $X=G$
    or $S$ represents  the conductance or the thermopower,  $X^\prime/X$
    denotes the ratio of the conductance or  
    thermopower after modifying
    the middle part of the system to its original value. Solid (dashed)
    curves are for $X=S$ ($X=G$) for a three-site model. The
    $\bullet$ ($\triangle$) points denote $X=S$ ($X=G$) for a 1D NNH
    model with 31 LSs. Note that for both models $S^\prime/S=1$. The parameters for the 31 LSs 1D NNH
    model are: $k_BT=$20, $\mu=0$, and $\xi=0.001$. The LSs
    are located at the sites of a 1D lattice with a periodicity of
    $0.008$. The energy is uniformly distributed in the energy window
    $(-E_{c}, E_{c})$ with $E_{c}=60$.                
    The Mott hopping
    length is $
    \sqrt{\xi/\rho k_BT}/2=0.004$. This value,  half of the 
    nearest-neighbor distance, makes all the hopping processes except the NN one 
    sufficiently small. By
    ``decreasing the conductances in the middle of the system'' we
    mean  decreasing the conductances of the connections among the 
    1st, 2nd, and 3rd LSs,
    while for the 31 LSs 1D NNH model it means decreasing the conductance of all
    bonds between the 11th and 21st LSs by a factor of $r_g$.}\label{fig2}
\end{figure}

To demonstrate the boundary effect in the NNH regime we study a simple model system
which consists of just  three LSs. Consider the situation where LSs 1 and 3 are strongly
coupled, by elastic transitions, to the lead continua on  the left and on the right,  respectively so that $f_1 =
f_L(\vep_1)$ and $f_3 = f_R(\vep_3)$. That is, the conductances
$G_{1L}$ and $G_{3R}$ are much larger than the other conductances. It
is also assumed that the tunneling conductances between  LS2 and
the leads are so small that the transport through the system is
dominated by the hopping path illustrated in Fig.~\ref{fig2}(a). The
condition for this to be true is analyzed in detail in
Appendix~\ref{ap:tun}. This model system can be realized in experiments by, say, a
serially coupled three-quantum-dots .\cite{hop-dot} For
concreteness, consider the situation when $\vep_1,\vep_3<\vep_2$. $U_2$ is then determined by  $I_{1\to 2}= I_{2\to 3}$ where
\bea
I_{1\to 2}&=& G_{12}(U_1 - U_2 + U_{12})\ , \nn\\
I_{2\to 3}&=& G_{23}(U_2 - U_3 - U_{23}) \ , \label{gx2}
\eea
and consequently
\begin{align}
U_2 = \frac{G_{12}(U_1+U_{12}) +
  G_{23}(U_3 +U_{23})}{G_{12}+G_{23}}\ , \label{x2}
  \end{align}
  and
  \begin{align}
I_e=
\frac{G_{12}G_{23}}{G_{12}+G_{23}} [U_3 +U_{23}-U_1-U_{12}] \ ,
\end{align}
where
$I_e = I_{1\to 2}$ is the total electric current.
Expressing the $U$'s as functions of the chemical potential and
temperature differences gives
\be
I_e = \frac{G_{12}G_{23}}{G_{12}+G_{23}}
[\frac{\delta\mu}{e} +
\frac{\ov{\vep}_{31}}{e}\frac{\delta
  T}{T}+\frac{\ome_{31}}{e}\frac{\Delta T}{T}]\  ,\label{rem}
\ee
where we have denoted $\ov{\vep}_{31}=(\vep_1+\vep_3)/2-\mu$
and
$\ome_{31}=\vep_3-\vep_1$. Using Eqs.~(\ref{is1}) and (\ref{is2}) one
finds
\be
\dot Q_L = -\frac{\vep_1-\mu}{e} I_e\ ,\quad \dot Q_R =
\frac{\vep_3-\mu}{e} I_e
\  .
\ee
Inserting these results into Eq. (\ref{trans}) yields the transport coefficients in the linear-response regime, 
\begin{align} 
& L_{1}=G\left(
\frac{\ov{\vep}_{31}}{e}\right),\     L_2 = G\left( \frac{\ome_{31}}{e}\right) , \ 
 K_e^0 = G\left(\frac{\ov{\vep}_{31}}{e}\right)^2,\  \nonumber\\
& L_{3} = G
\left(\frac{\ov{\vep}_{31}}{e}\right)\left(\frac{\ome_{31}}{e}\right),\
  K_{pe}=G\left(\frac{\ome_{31}}{e}\right)^2 ,
\end{align}
with 
\be
G=\frac{G_{12}G_{23}}{G_{12}+G_{23}} \ ,\label{g123}
\ee
and confirms the Onsager reciprocity relations.
In Appendix~\ref{ap:prob} we  reproduce these results by a
probabilistic analysis.

Remarkably, the thermopower $S=L_1/(TG)$ as well as the coefficients  $L_2/G$, $L_3/G$,
$K_e^0/G$, and $K_{pe}/G$ depend all only on $\vep_1$ and $\vep_3$, i.e.,
the energies of the LSs at the boundaries. The  site energy of the central level does 
{\em not} affect these quantities. 
{\em The thermoelectric properties are completely determined at the
boundaries}. On the contrary, the bulk (in this simple example, the central level) can  affect the conductance of
the system. 
Clearly, we can replace the middle site by a more complicated 
construction. As long as it is  coupled to the boundary sites in the same way, the properties of this mid-system do not matter for the above-mentioned transport coefficients! 

To illustrate the 3-site case, we have numerically computed the conductance $G$ and the
thermopower $S$ of a three-site system as a function of the decrease of
the conductance of the bonds (1,2) and (2,3). Namely,  we have determined 
the conductance and the thermopower of the system when $G_{12}\to
G_{12}/r_g$ and $G_{23}\to G_{23}/r_g$. We 
plot the ratio of the new conductance $G^\prime$ to the original one
$G$ as well as the ratio $S^\prime/S$ as  functions of the scale factor  $r_g$ in
Fig.~\ref{fig2}(b). It is seen that although the conductance decreases
significantly with increasing $r_g$, the thermopower remains unchanged, $S^\prime/S= 1$. More complicated models will be discussed below.

\subsection{Longer 1D NNH systems}

\label{LONGER}

We now extend the discussion to longer 1D hopping systems. Nearest-neighbor hopping   in
a chain of LSs is accomplished  via electron transits into the left (right) lead only
through the leftmost (rightmost) LS, having energies $\vep_l$ ($\vep_r$). Therefore from Eqs.~(\ref{is1})
and (\ref{is2}) 
\be
\dot Q_L = -\frac{\vep_{\ell}-\mu}{e} I_e,\quad \dot Q_R =
\frac{\vep_r-\mu}{e} I_e\  ,
\ee
and one readily finds
\be
I_Q^e = \frac{\ov{\vep}_{r\ell}}{e} I_e\ , \quad I_Q^{pe} =
\frac{\ome_{r\ell}}{e} I_e \ ,\label{i-rel} 
\ee
with 
$\ov{\vep}_{r\ell}=(\vep_{\ell}+\vep_r)/2-\mu$
and
$\ome_{r\ell}=\vep_r - \vep_\ell$. Interestingly enough,  the thermoelectric
properties can be deduced without solving the resistor
network. For example, when 
$\delta\mu\ne 0$ and
$\delta T=\Delta T=0$, one has $I_e=G\delta\mu/e$
and then by 
Eq.~(\ref{i-rel}) 
$I_Q^e=(\ov{\vep}_{r\ell}/e)I_e=G
\ov{\vep}_{r\ell}\delta\mu/e^{2}$
and $I_Q^{pe}=(\ome_{r\ell}/e)I_e=G
\ome_{r\ell}\delta\mu/e^{2}$.
Therefore
\be
L_{1}=G \left(\frac{\ov{\vep}_{r\ell}}{e}\right) ,\quad     L_2 = G\left(
\frac{\ome_{r\ell}}{e}\right)\   . \label{re1}
\ee
Analyzing the situations when $\delta T\ne 0$ and
$\delta\mu=\Delta T=0$ and when $\Delta T\ne 0$ and $\delta\mu=\delta
T=0$ and exploiting the Onsager reciprocity relations one  obtains
\begin{align} 
& K_e^0 = G\left(\frac{\ov{\vep}_{r\ell}}{e}\right)^2 ,\  
 L_{3} =
 G\left(\frac{\ov{\vep}_{r\ell}}{e}\right)\left(\frac{\ome_{r\ell}}{e}\right), \nonumber\\
&  K_{pe}=G\left(\frac{\ome_{r\ell}}{e}\right)^2   \label{re2} .
\end{align}
Again the thermoelectric properties are completely determined at the
boundaries. This is clearly manifested in Fig.~\ref{fig2}(b) for a NNH
system with 31 LSs. It is seen that the thermopower is immune to the
change in the middle part of the system while the conductance is
considerably affected. Remarkably, this also implies  that the
thermopower (and other thermoelectric coefficients) is {\em   finite
  and random} (as long as $\vep_{\ell}$ and $\vep_r$ are finite and
random) regardless of the bulk properties. This even
persists, in general, for very long systems where the particle-hole asymmetry is
negligible, as long as the edge sites have definite energies (and their sum 
does not happen to vanish exactly). Therefore the particle-hole symmetry  may no longer
dictate a zero sample-specific thermopower in the macroscopic limit in 1D NNH systems.

\section{ Simplified 2D NNH systems}

\label{SIMPLIFIED}

\begin{figure}[htb]
  \centering \includegraphics[height=3cm]{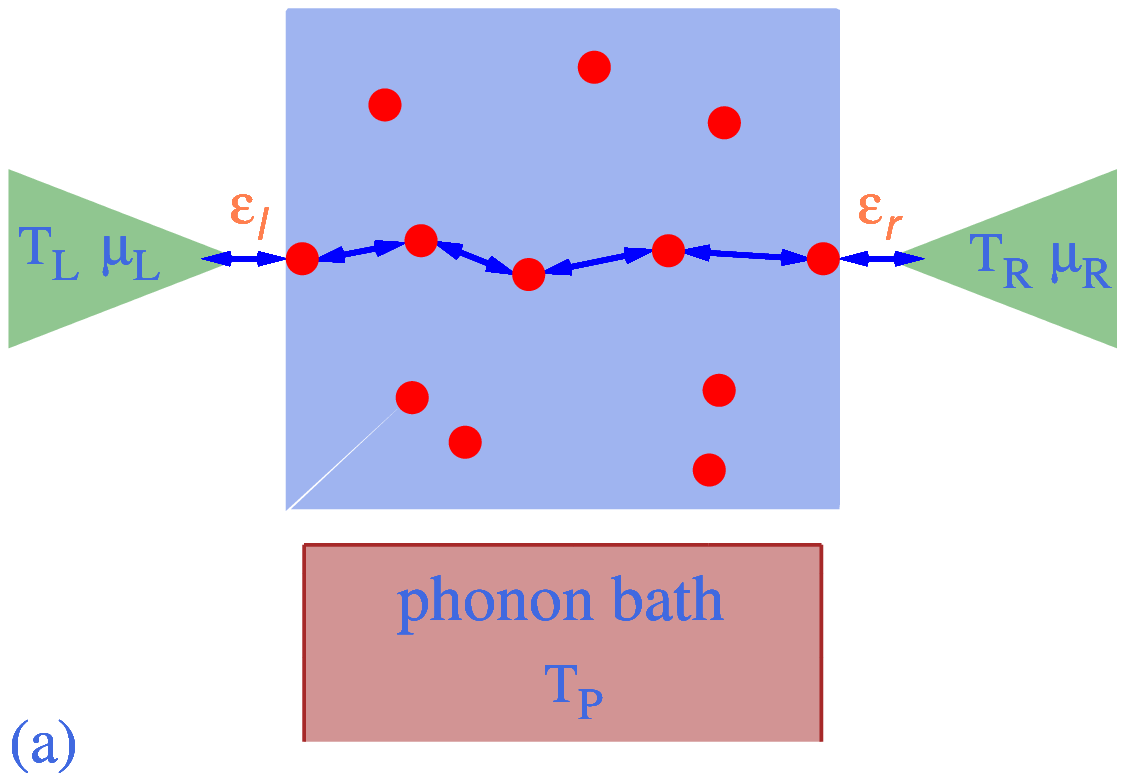}\includegraphics[height=3cm]{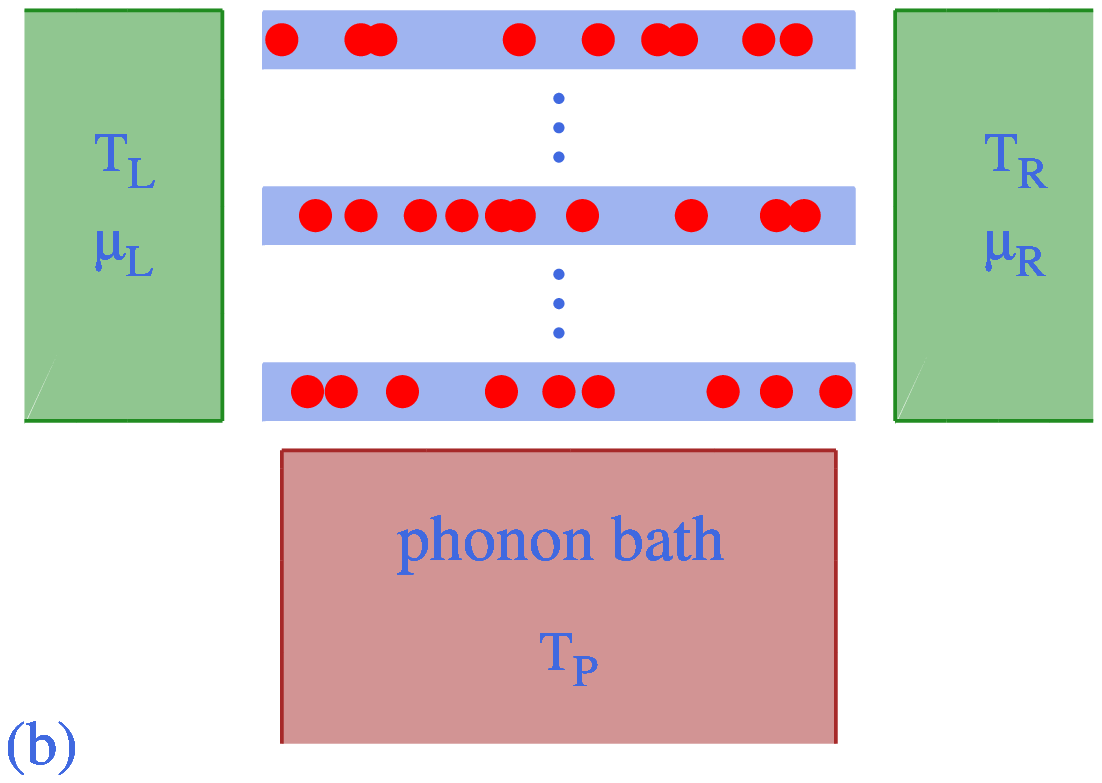}
 \caption{(Color online)  Three-terminal hopping transport in a 2D NNH
    system. (a): Illustration of the situation when the two  electronic
    terminals are geometrically sharp. Each terminal is strongly
    coupled only with a single LS with energy $\vep_{\ell}$  and
    $\vep_r$ on the left and right, respectively. A possible hopping path is
    illustrated in the figure by the arrows. (b): A 2D system consisting of a series of 
    parallel hopping chains between the two electronic leads. The
    hopping between different chains is negligible.} \label{fig5}
\end{figure}

We start by studying a situation where the boundary effect fully dominates
thermoelectric properties: When the electronic leads are geometrically 
sharp (as with a high resolution STM configuration) and each of them is coupled strongly only with a single LS as
illustrated in Fig.~\ref{fig5}(a). Specifically the left lead is coupled with a LS
of energy $\vep_{\ell}$ while the right one with a LS of energy
$\vep_r$. In this way the relation between the two heat currents
($I_Q^e$ and $I_Q^{pe}$) and the electric current $I_e$ is given  again by
Eqs.~(\ref{i-rel}). Following the same logic as that applied for 1D NNH
systems one  again obtains Eqs.~(\ref{re2}). Therefore the
thermoelectric properties are completely determined by the boundary
LSs (i.e., the LSs coupled strongly with the two electronic leads)
in this case as well.

\begin{figure}[htb]
  \centering \includegraphics[height=3.5cm]{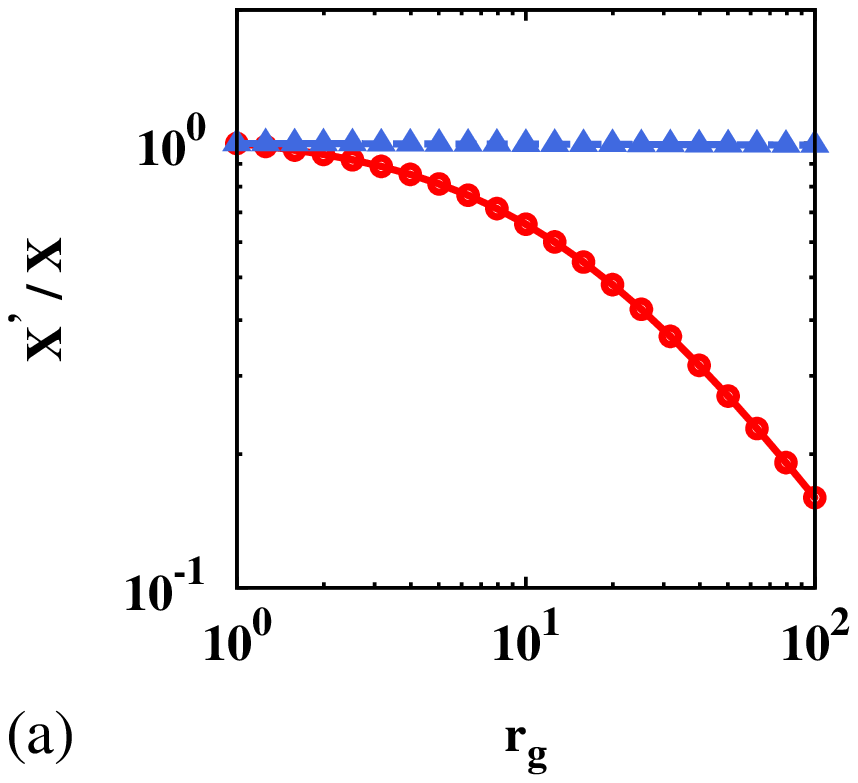}\includegraphics[height=3.5cm]{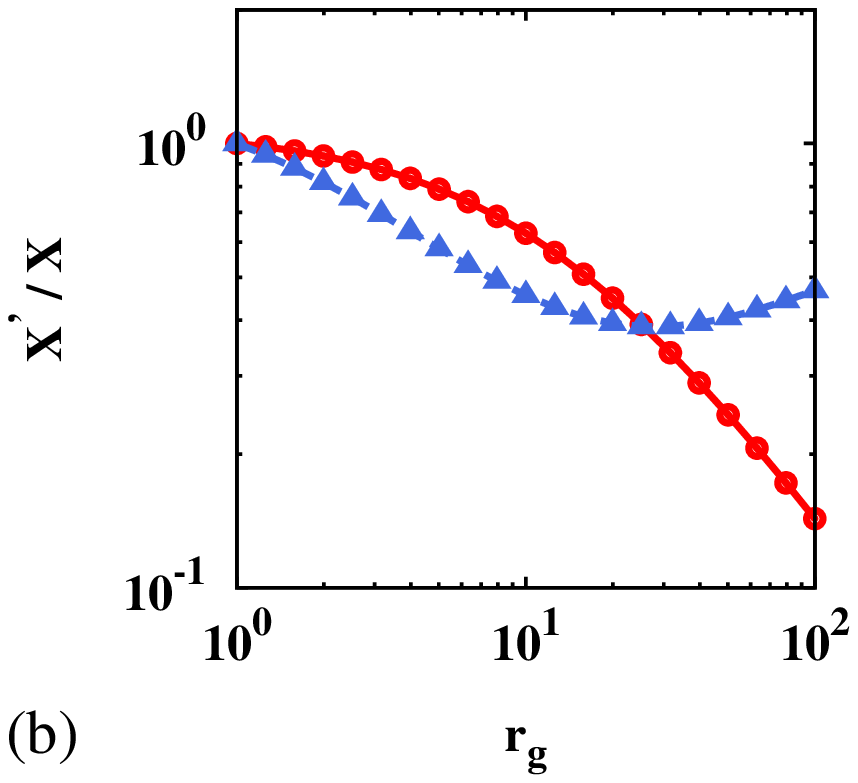}
  \caption{(Color online) 
   The effect of decreasing the conductances in the
    middle of the system by a factor of $r_g$ on the conductance and
    thermopower,  for a 2D NNH single realization. $X=G$ or $S$ represents the
    conductance or the thermopower, and $X^\prime/X$ denotes the ratio of
    the conductance or thermopower after changing the middle to the
    original value. The curves with $\triangle$ are for   the thermopower
    while the ones with $\bullet$ are for the conductance. The
    parameters are: $k_BT=$20, $\mu=0$, $\xi=0.01$, $E_{c}=20$, and
    $\rho=40$. The Mott hopping length is      $[\xi/(\rho
    k_BT)]^{1/3}/2=0.012$. The LSs are put on a 2D lattice with 
    periodicity $0.025$. The number of lattice 
    sites perpendicular to the transport direction is $N_y=20$. For
    the longer system (a) the number of sites along the transport
    direction is $N_x=31$. For the shorter system (b) $N_x=15$. Each
    lattice site is denoted by two indices $(i_x, i_y)$ with     
    $i_x\in[-(N_x-1)/2, (N_x-1)/2]$    
   and $i_y\in[1,
    N_y]$. The term ``conductance in the middle'' 
    stands for the conductance between two LSs $(i_{x1},i_{y1})$ and
    $(i_{x2},i_{y2})$ where $i_{x1}\in [-5,5]$ and $i_{x2}\in [-5,5]$.} \label{fig8}  
\end{figure}

Next we consider another special type of 2D systems that consist of
parallel 1D hopping chains where there is no transport between
different chains as sketched in Fig.~\ref{fig5}(b). The heat and electric
currents are given by
\bea
I_e = \sum_k I_e^{(k)} ,\quad I_Q^e = \sum_k \frac{\ov{\vep}_{r\ell}^{(k)}}{e}I_e^{(k)} ,
\quad I_Q^{pe} = \sum_k \frac{\ome_{r\ell}^{(k)}}{e} I_e^{(k)} , \nn
\eea
where the superscript $k$ denotes the $k$-th chain. One then has
\begin{align}
&\hspace{-0.1cm} L_{1}=G
\ave{\frac{\ov{\vep}_{r\ell}}{e}},\     L_2 = G \ave{\frac{\ome_{r\ell}}{e}} , \
 K_e^0 = G\ave{\left(\frac{\ov{\vep}_{r\ell}}{e}\right)^2},\  \nonumber\\
&\hspace{-0.1cm} L_{3} = G\ave{\left(\frac{\ov{\vep}_{r\ell}}{e}\right)\left(\frac{\ome_{r\ell}}{e}\right)},\
  K_{pe}=G\ave{\left(\frac{\ome_{r\ell}}{e}\right)^2} \  ,\label{23}
\end{align}
where $G=\sum_k G_k$ with $G_k$ being the conductance of the $k$-th
chain and 
\be
\ave{...} = \frac{\sum_k G_k ...}{\sum_k G_k} \ .\label{Gk}
\ee
Altering the central region will modify the conductances $G_k$. Due to
the random nature of the system this modification varies among
different chains and  changes  the averaging in Eq.~(\ref{Gk}). 
Although $\ov{\vep}_{r\ell}$ and $\ome_{r\ell}$ for
each chain do not change, the averaged values in Eqs. (\ref{23}) do.
Therefore the thermoelectric properties also depend on the bulk in
this type of 2D systems. This is also true for other types of 2D
NNH systems where the backbone consists of parallel hopping
paths. Even when there are connections between those parallel paths,
if the current distribution at the boundaries can be considerably
affected by the bulk, the bulk effect cannot be ignored. However
for sufficiently long 2D systems the current distribution at the
boundaries should not be affected by the far away bulk. In general there
is a correlation length $L_{co}$ in hopping systems beyond which the
spatial current distributions are uncorrelated.\cite{Shklovski} In 1D
NNH systems $L_{co}$ is on the order of the distance between the adjacent
LSs. In 2D NNH $L_{co}$ can be much larger but still finite. In
Fig.~\ref{fig8} we show how the conductance and the thermopower are
affected by  changes in the middle part of   a long and a short 2D NNH
system. In the long system the thermopower is almost unchanged whereas
in the short one it is significantly modified (but, possibly much less than the conductance, which can be, for example, affected exponentially). $L_{co}$ for the chosen
parameters is estimated as $\sim 6$ times  the distance between the
adjacent LSs.


We now discuss the macroscopic limit for these  2D hopping systems. For sufficiently long systems, 
the fluctuation in the
conductance of each chain becomes very small.\cite{1dvrh0} The averages
in Eqs. (\ref{23})  become the averages over the energies at the
boundaries. If the energies of LSs at the boundaries obey the same (sufficiently broad)
distribution as the bulk ones,  then 
\be
\ave{...} \simeq \frac{\int d\vep\rho(\vep) ...}{\int d\vep\rho(\vep)} \ ,\label{ave}
\ee
with $\rho(\vep)$ being the density of states of the involved
LSs. 
When the density of states is symmetric with respect to the
chemical potential, one finds 
\begin{align}
& L_{1}= 0,\     L_2 = 0 ,\  L_{3} = 0 ,\
 \nonumber\\
& K_e^0 = \frac{1}{2}G\frac{\ave{\vep^2}}{e^2},\ 
K_{pe}=2G\frac{\ave{\vep^2}}{e^2} \  ,\label{26}
\end{align}
with the average given by Eq.~(\ref{ave}). The particle-hole symmetry
indeed leads to zero thermopowers in the bulk limit. In deriving Eqs. (\ref{26})  we have used  
$\ave{\vep_\ell}=\ave{\vep_r}$ and $\ave{\vep_\ell^2}=\ave{\vep_r^2}=\ave{\vep^2}$, and have taken into account the fact
that for very long 1D systems there should be no
correlation between $\vep_{\ell}$ and $\vep_{r}$. When $\ave{\vep_\ell}=\ave{\vep_r}$
then $L_2=L_3=0$ in the bulk limit even  when particle-hole
symmetry is broken, such that $L_1\ne 0$. In fact $S_p=L_2/(TG)$ and $L_3$ have
nothing to do with the particle-hole asymmetry but are related to the
(left-to-right) inversion asymmetry in the sense that $S_p\propto
\ave{\vep_r}-\ave{\vep_\ell}$ and $L_3\propto
\ave{\vep_r^2}-\ave{\vep_\ell^2}$.

\section{Boundary effect in 1D VRH systems}

\label{VRHSEC}

Here we will not assume that only a single LS is strongly coupled to
each lead. 
Quite generally, all  LSs located within
a distance from the lead smaller than or comparable to the Mott hopping distance $R_M$ can
be considerably coupled to that lead, with the conductance of the connection given  by Eq. (\ref{gijl}). Other LSs, which are not coupled directly
with the leads have much lower conductances, due to the exponential decay of $G_{iL}$
and $G_{iR}$ with the distance $r_{iL}$ and $r_{iR}$. This implies that  the
boundary effect is somewhat weakened. To study this situation, we consider 
sufficiently long 1D VRH systems whose length $L$ is much larger than $R_{M}$ and denote by ``boundaries" the
regions that are within a distance of a few $R_M$'s from the leads. We
shall 
find that the boundary effect on the VRH thermopower is still important.

Specifically for VRH systems the
current flowing into each lead comes mainly from the LSs in the boundary
regions. The summations in Eqs.~(\ref{is1}) and (\ref{is2})
are then reduced to summations over those LSs. Accordingly, the
thermopowers can be written as
\be
S = \frac{1}{eT}\left(\frac{\ave{\vep_r}+\ave{\vep_\ell}}{2}-\mu\right),\quad S_p =
\frac{1}{eT} \left(\ave{\vep_r}-\ave{\vep_\ell}\right)\ ,
\ee
where 
\bea
\ave{\vep_r}=\frac{\sum_{i}^\prime\vep_i I_{i\to R}}{\sum_i^\prime
  I_{i\to R}} , \quad \ave{\vep_\ell} = \frac{\sum_{i}^{\prime\prime} \vep_i I_{i\to L}}{\sum_i^{\prime\prime}
  I_{i\to L}}\ , \label{bb}
\eea
with $\sum_i^\prime$ ($\sum_i^{\prime\prime}$) being restricted to the
LSs in the right (left) boundary region. In 1D NNH systems the
summation is restricted to a single LS that is coupled strongly with each
lead, while in 1D VRH systems there are more than one such LSs.  Nevertheless,
whenever the number of LSs involved in each summation is not too large the thermopowers will be finite and    will fluctuate 
regardless of the particle-hole symmetry in the bulk. The boundary part of the backbone
picture is drawn in Fig.~\ref{fig3}(a).  
Similar to NNH 2D systems,  the weights of the various $i$'s
In Eqs. \ref{bb} (e,g, $ I_{i\to L} $) does depend on the bulk. Therefore in a given sample,
there will again exist some limited dependence of the thermopowers on the bulk. However, this will be averaged out in an ensemble of many realizations of the sample.

\subsection{Thermopowers}

\label{VRHS}

\begin{figure}[htb]
  \centering \includegraphics[height=3.85cm]{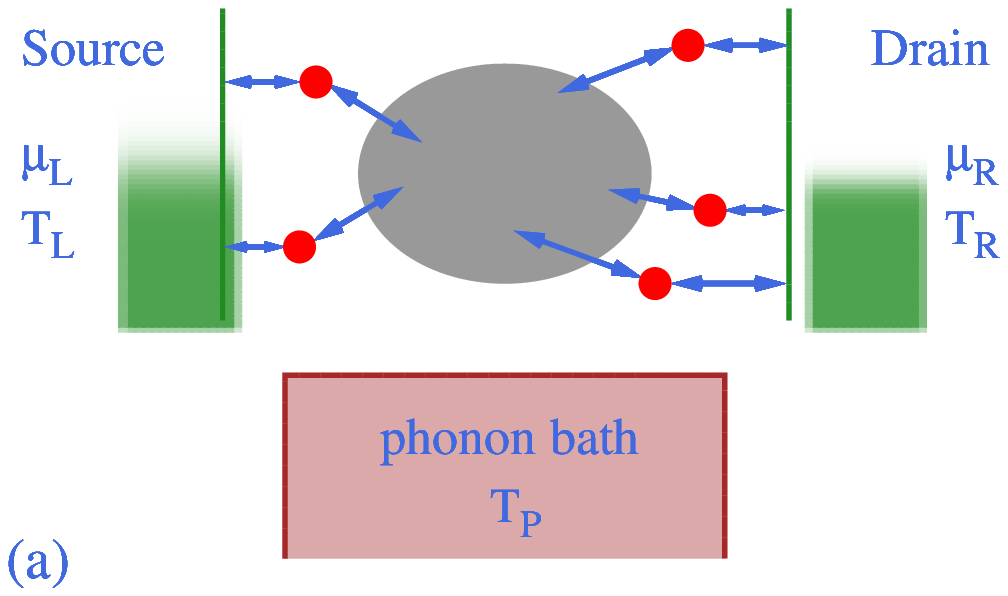}
  \centering \includegraphics[height=4cm]{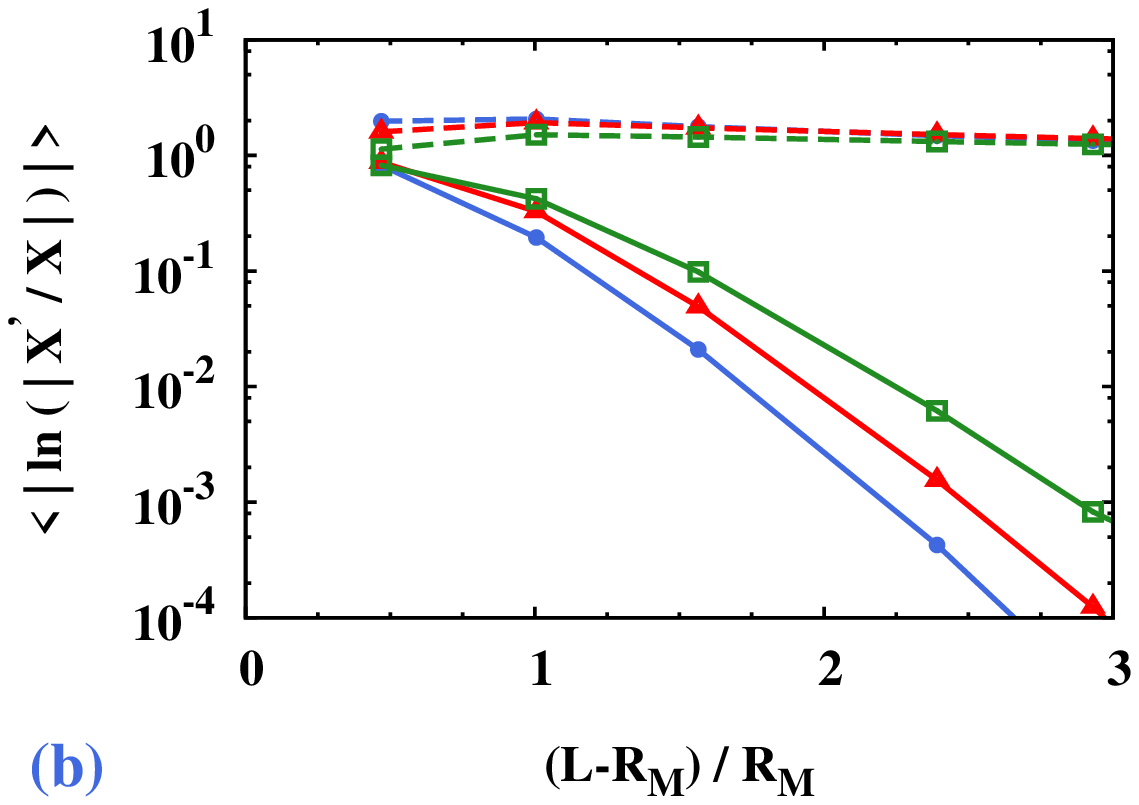}
  \centering \includegraphics[height=4cm]{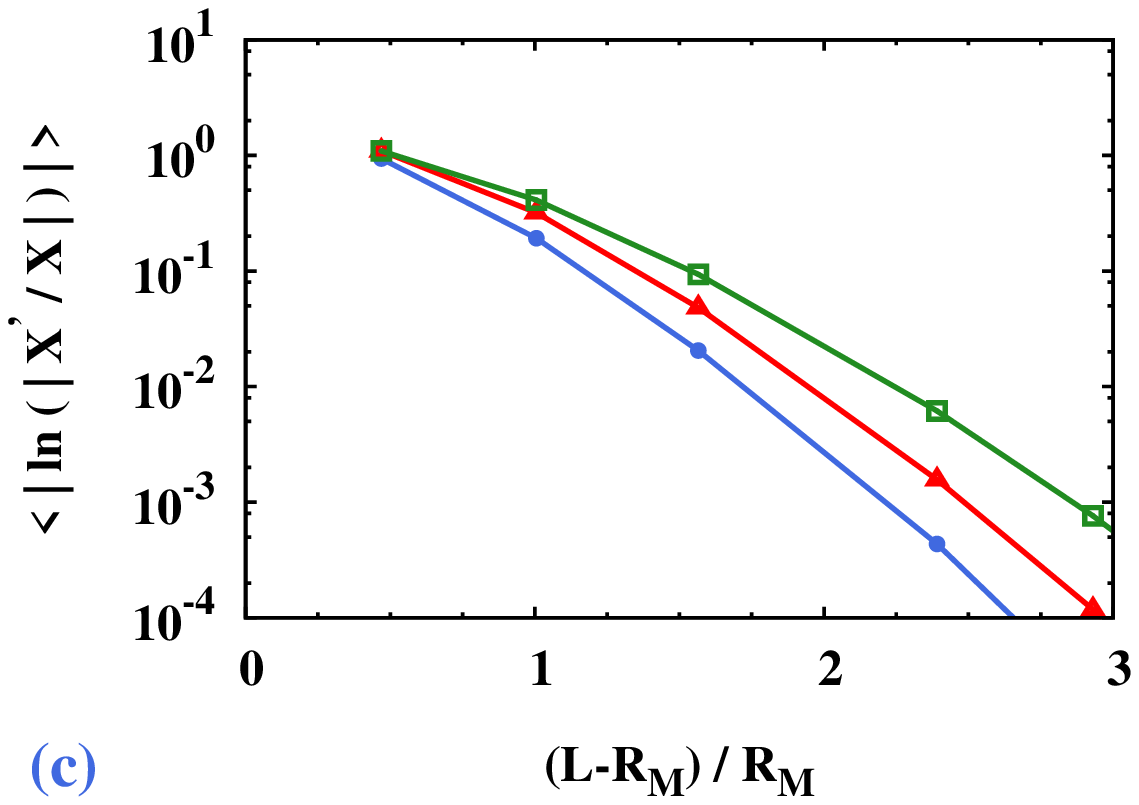}
  \caption{(Color online) 1D VRH. (a): The boundary part of the \Red{hopping} backbone, generalizing Fig. \ref{fig2}a. The gray region depicts the main part of the
    backbone. The connections (arrows) to the leads are through the LSs (red dots) near the boundaries.
    (b) and (c):
    Length $(L-R_M)/R_M$ dependence of the average
    $\ave{|\ln(|X^\prime/X|)|}$. In (b): the conductance $X=G$ (dashed
    curves) and the thermopower $X=S$ (solid curves);  In (c): the
    three-terminal thermopower $X=S_p$. Results given for three different
    temperatures $k_BT=$15 ($\bullet$), 30 ($\triangle$), and 60
    ($\square$). $r_g=100$, $\mu=0$, $\xi=0.1$, and $\rho=0.03$, $k_B T_0=333$, $E_{c}=424$,
    600, and 849 for the three temperatures  respectively. Along each
    curve  from left to right the number of  LSs for the first,
    second, ..., fifth data point are $N=30$, 50,
    70, 100, and 120 respectively. The Mott distances are $R_M=0.4, 0.49, 0.54, 0.58, 0.60
    (0.28, 0.35, 0.38, 0.41, 0.42)$ and the corresponding Mott energies are $E_M=120, 147, 161, 174, 180
    (170, 208, 227, 246, 254)$ for the five points at $k_BT=15 (30)$ respectively.
     $R_M$ ($E_M$) at $k_BT=60$ is half (two times) of that at $k_BT=15$ for the five points
     respectively. The average nearest neighbor distances  $2L/N$ for the three $T$'s are 0.039, 0.028,
    and 0.02 respectively. The results  are averaged over $10^6$
    random configurations. The  curves in (b) and (c) are guide to the eye.}\label{fig3}
\end{figure}

We now turn to the thermoelectric properties of VRH systems in
1D. The conductance of a 1D  system is mainly suppressed by the
``breaks'',\cite{1dvrh0,1dvrh} rendering   the Mott VRH picture   not
entirely  applicable. At low temperatures the
characteristic conductance of a 1D VRH system of length $2L$
is\cite{1dvrh0,1dvrh} 
\be
G_{1D}=G_0 e^{-\eta} \ ,
\ee
where
\bea
\eta = \left(\frac{T_M}{T}\right)^{1/2}, \quad
T_M = 2 T_0 \ln\left(\frac{2\sqrt{\nu}L}{\xi}\right)\ , \label{tm}
\eea
with $k_BT_0=(\rho\xi)^{-1}$ and $\rho$ denoting the density of
(localized) states. $\nu$ is the solution of 
\cite{1dvrh0,1dvrh} 
\be
\nu = \frac{2T}{T_0}\ln\left(\frac{2\sqrt{\nu}L}{\xi}\right)\ .\label{nu}
\ee
The current  mainly flows in  the backbone of the resistor
network which mostly consists of connections with conductance higher
than or comparable with $G_{1D}$.\cite{1dvrh0,1dvrh,1dvrh2} The
typical hopping length and energy are $R_M = \eta \xi/2$ and
$E_M = \eta k_B T$. Below we use the
energy scale 
\be
E_0=k_B\sqrt{T_0T}\ ,\label {e0}
\ee 
which does not depend on the system length $2L$. 
For example, $T$ in units of this scale will be seen to be relevant for the thermopower fluctuations, see Fig. \ref{fig4}(b).

The boundary effect is detected 
by comparing the thermopower
$S$ (and other thermoelectric coefficients) of a random system and
that of the same configuration but with the central region
modified, $S'$. Specifically,  we apply the following modification: $G_{ij}
\to 10^{-2}G_{ij}$, if both $i$ and $j$ are in the central
part. 
The boundary
effect is monitored by $|\ln(|S^\prime/S|)|$. Concomitantly we 
compute the conductances of the original and the modified systems, $G$
and $G^\prime$, and monitor the change via 
$|\ln(G^\prime/G)|$. If $|\ln(|S^\prime/S|)|$ is very small (i.e.,
$S^\prime$ is almost the same as $S$) then the thermopower is
insensitive to the bulk and the boundary effect dominates. We 
model the localized electron system by a number of LSs located at  random
positions and and having  energies which are 
uniformly distributed in the ranges
$(-L,L)$ and $(-E_{c},E_{c})$, respectively. The central region is
taken as $x\in (-R_M, R_M)$. The linear-response
transport coefficients are computed using the method described in
Appendix~\ref{app:num}. Note that for the numerics we use dimensionless energy 
and temperature, with $k_B = 1$. For a given system, the appropriate energy unit can be introduced.

The averages of the two quantities over $10^6$ random configurations
are plotted in Fig.~\ref{fig3}(b) and (c). It is seen that the conductance is
considerably modified, $\ave{|\ln(G^\prime/G)|}>1$. In contrast the
change in thermopower is much smaller, especially when the distance
between the central region and the boundary $L-R_M$ exceeds the
hopping length $R_M$. $\ave{|\ln(|S^\prime/S|)|}$
decays rapidly with the distance $L-R_M$ and soon becomes
negligible.

For the choice of the central region adopted in the figure (from $-R_M$
to $R_M$), the change of the conductance is not dramatic, e.g., $G'/G \cong 1/4$ for the
last point (N=120). But if the central region is taken to be from $-2R_M$ to $2R_M$
then $G'/G \cong 1/22$. The relative changes in thermopower  in the former and latter cases
are, however, no larger than $8\times 10^{-4}$ and $2\times 10^{-2}$ respectively.

The three-terminal thermopower $S_p$, shown  in Fig.~\ref{fig3}(c, behaves 
similarly. Therefore the probability that the LSs far
away from the boundaries can affect the average thermopowers is very
small. This also indicates that the correlation length, $L_{co}$, giving the scale 
over which a local change in the network influences the conducting path, in 1D VRH system is
only a few hopping lengths.

\begin{figure}[htb]
  \centering \includegraphics[height=3.5cm]{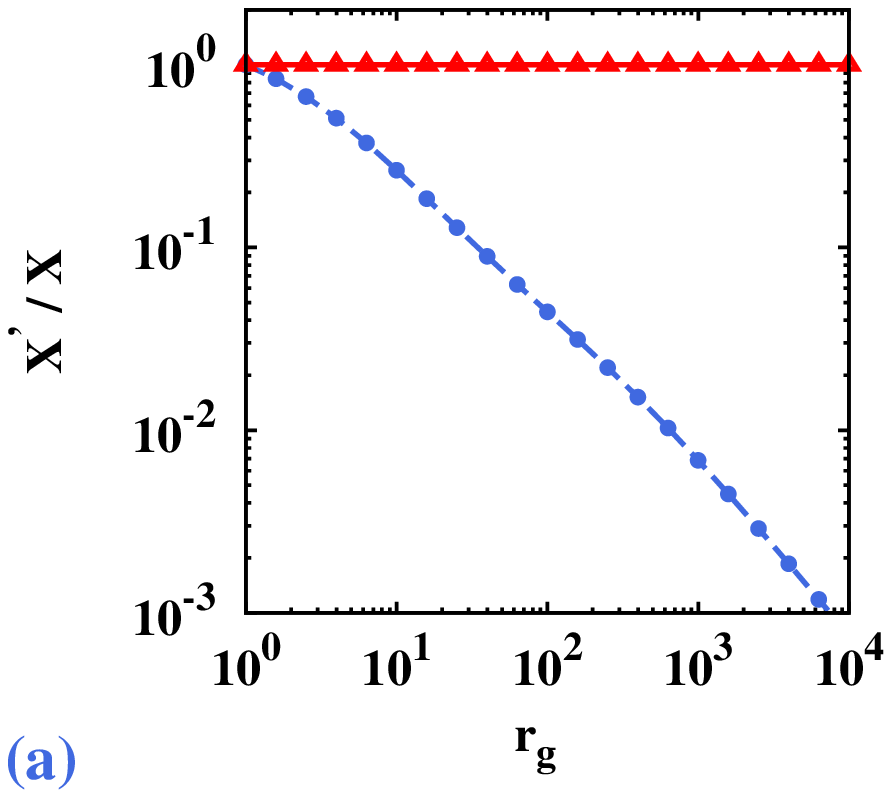}\includegraphics[height=3.5cm]{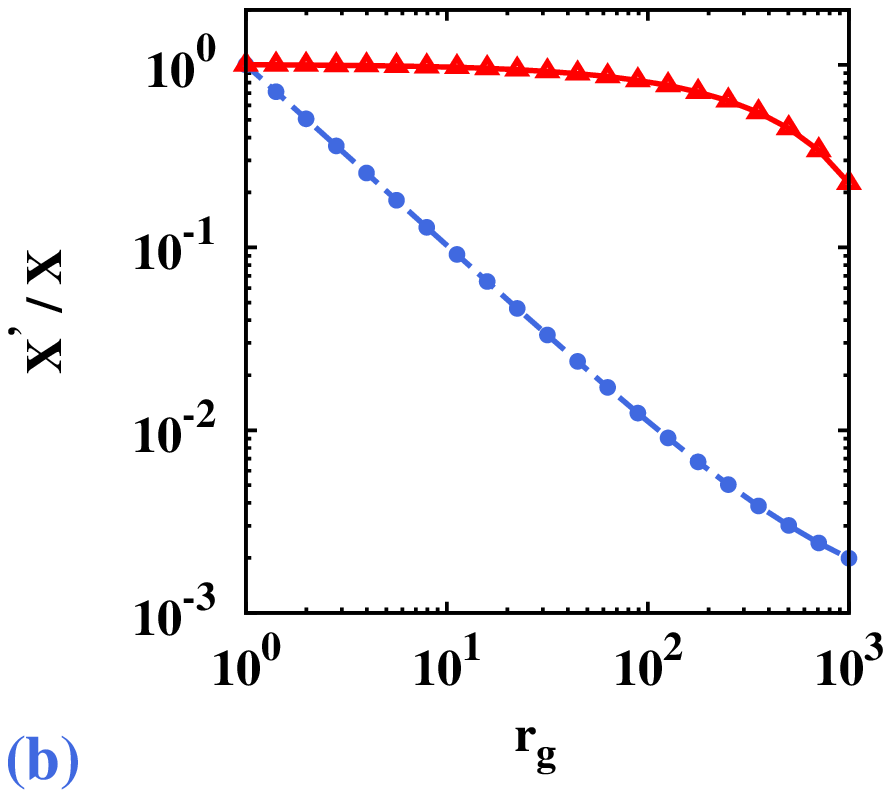}
  \caption{(Color online) 
   The effect of  decreasing the conductances in the
    middle part of a random 1D VRH system,  by a factor of $r_g$,  on the conductance and the
    thermopower. $X=G$ or $S$, with 
    $X^\prime/X$ denoting the ratio of
    the conductance or thermopower after changing the middle, to the
    original value. The curves with $\triangle$ depict the thermopower while the
    ones with $\bullet$ depict the conductance. The parameters are: $k_BT=$15,
    $\mu=0$, $\xi=0.1$, $\rho=0.03$, $k_B T_0=333$ and $E_{c}=424$. The average distance
    between adjacent LSs is $0.039$. For the longer (a) system ($N=400$,
    $2L=22R_M$) the resistances in the region $(-2R_M,2R_M)$ are
    increased by a factor of $r_g$. For the shorter (b) system ($N=40$, $2L= 3.5R_M$) the
    resistances in the region $(-R_M,R_M)$ are increased in the same way.} \label{fig7} 
\end{figure}


The question naturally arises, what happens in a specific sample?
We find that there again  the  boundary effect can  be dominant. In Fig.~(\ref{fig7}) we plot the change of the
thermopower and the  conductance for two systems as a function of the
increase of the resistance in the middle part of the sample. For the longer system it is
seen that the thermopower is unaffected while the conductance decreases by
almost three orders of magnitude. This is the situation when the distance between the boundary and
the middle exceeds the correlation length $L_{co}$  
so that the bulk
affects the thermopower negligibly. However for a shorter system
a change in the  central part can affect both the thermopower and the
conductance. Nevertheless the change in the conductance is still much
more significant than that in the thermopower.

Interestingly enough, our analysis points out that the thermopower has unexpectedly large sample-to-sample fluctuations even for very large samples. This is very different from the vanishing of the VRH conductance fluctuations for increasing-length samples.
No matter whether the bulk effect is important or not, as long as the
number of the LSs involved in the summation in Eqs.~(\ref{bb}) is
finite, the thermopowers  have a {\em finite and fluctuating}
value due to insufficient averaging. For 1D VRH, the LSs involved in the averaging, i.e., those with  $r_{iL}<R_{M}$ (or $r_{iR}<R_M$) and $|E_i|<E_M$ are typically just a few. \cite{AHL} Thus the fluctuations of the thermopower can be rather
large. To check this, we computed the variance of the thermopower 
as a function of the length of the
system. The results are shown  in Fig.~\ref{fig4}(a). Indeed the variance of the thermopower
remains considerably large and {\em attains a constant value} for
very long systems. The variance of the thermopower in very long systems
increases with increasing temperature (decreasing $R_M$). The appearance of a ``break''  (whose probability is exponentially small anyway) should not modify
the hopping  energy window considerably. If the break is sufficiently far from the boundary it should not affect the current distribution among the boundary LSs. Hence the break mechanism has
negligible effect on hopping thermopower although it greatly modifies
the hopping conductance.\cite{note6} As a result, the variance of the
thermopower ${\rm Var}(eST)=\ave{(eST-\ave{eST})^2}$ 
does {\em not} depend on the length of the system in the limit $2L\to
\infty$. That is, it 
becomes a
constant although $\ave{S}=0$ for systems with particle-hole symmetry
when $2L\to \infty$. In contrast, for the conductance, $\ave{\ln
  G}\to -\infty$ and ${\rm Var}(\ln G)=\ave{(\ln G-\ave{\ln G})^2}\to
0$ when $2L\to \infty$,\cite{1dvrh0} as the conductance is determined
by the bulk and is significantly affected by the break mechanism. 

We also
computed the probability distribution function of the thermopower for
a specific set of parameters and plotted it in Fig.~\ref{fig4}(b). It
is seen that the thermopower is mostly distributed in the range of
$-E_0<eST<E_0$. The probability distribution function is {\em not} a 
normal distribution.\cite{note5} Rather, it has exponential
tails, 
$\sim \exp[-C|eS|T/E_0]$,
at large $|S|$ with $C$ being
a constant depending on the parameters of the system. The exponential
tails should come from the fact that LSs with high energies have
exponentially small probabilities to be part of the backbone at the
boundaries because the resistance between such sites and the lead is exponentially
large. In the inset of Fig.~\ref{fig4}(b) we also show how the
thermopower evolves as  a function of the system length for two random
configurations. By increasing the system length there is an increasing probability to have, for example,   a poorly conducting piece
in the bulk of the system. It is seen that the
thermopower saturates with large system length since  the boundary effect
is dominant. Meanwhile the different thermopowers for the two
configurations vividly indicate the fluctuations of the thermopower even
in very long systems. Relatively large mesoscopic (sample to sample) fluctuations in the low-temperature thermopower
have been found also in the weak-disorder regime. \cite{Anis}

\begin{figure}[htb]
  \centering \includegraphics[height=4cm]{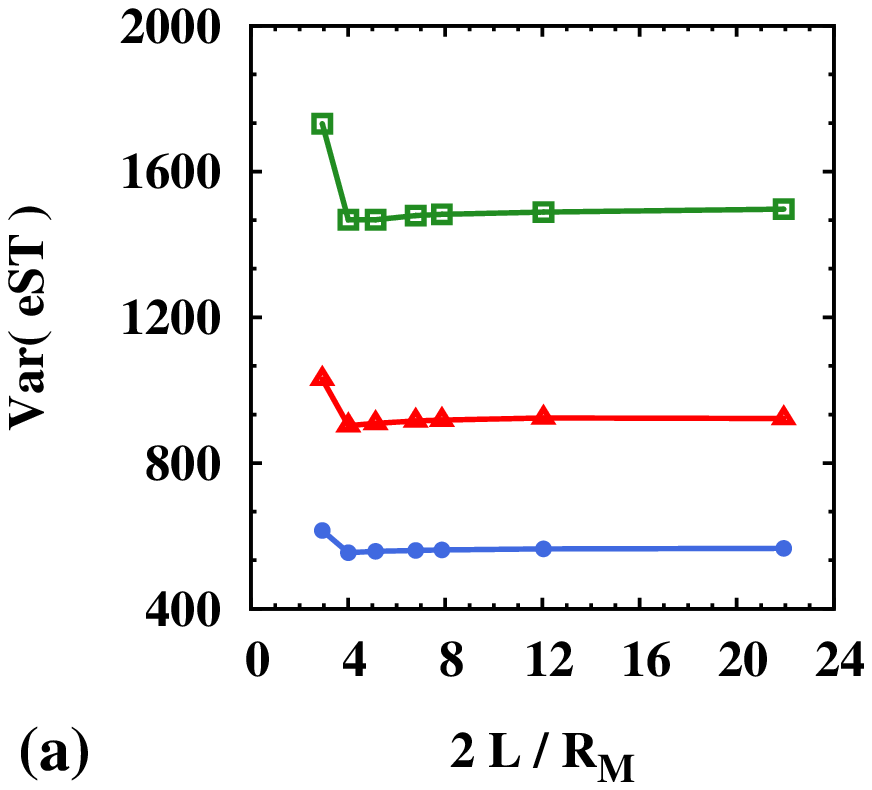}\includegraphics[height=4cm]{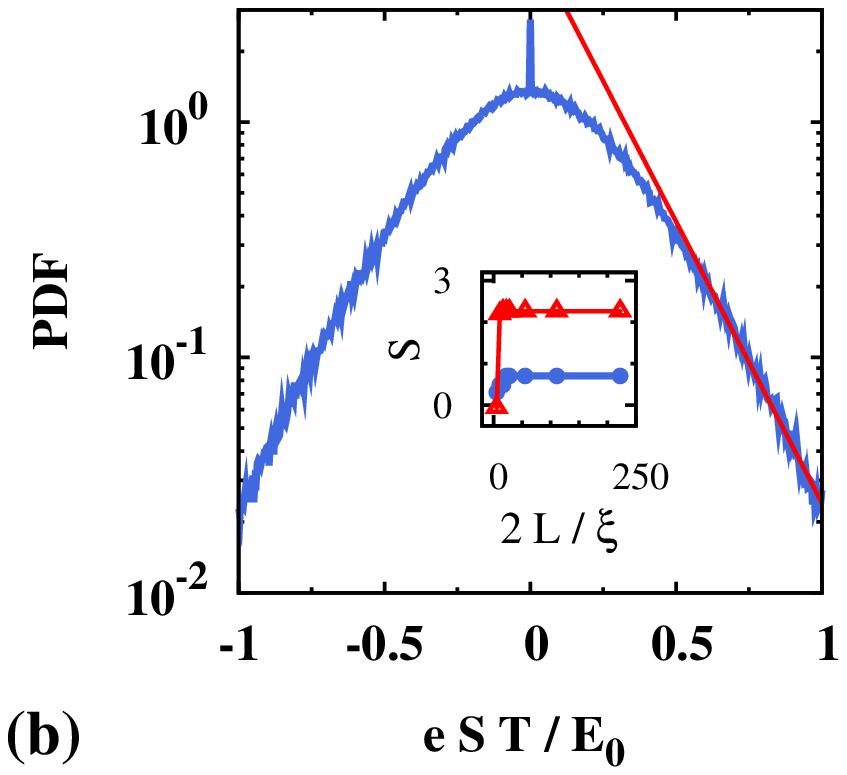}
  \caption{(Color online) Thermopower in1D VRH. (a): The variance of the thermopower ${\rm
      Var}(eST)=\ave{(eST-\ave{eST})^2}$ as a function of the length of
    the system $2L/R_M$ at $k_BT=$15 ($\bullet$), 30 ($\triangle$), and 60
    ($\square$). Along each curve  from left to right the number of LSs
    for the first, second, ..., seventh data point are $N=30$, 50, 70,
    100, 120, 200, and 400  respectively. The other parameters are the
    same as in Fig.~\ref{fig3}. (b): Probability distribution
    function (PDF) of the thermopower for 1D VRH systems, where
    $k_BT=15$, $\mu=0$, $\xi=0.1$, $\rho=0.03$, $E_{c}=424$, and $N=120$. The
    average distance is $2L/N=0.039$. The results are obtained from
    $10^6$ random configurations. The straight red line is an
    exponential fit to the tail, 
    $\sim \exp(-C|eS|T/E_0)$    with $C\simeq 5.5$. In the inset we show how the thermopower
    evolves as a function of  the system length for two random
    configurations. The parameters are: $k_BT=$30, $\mu=0$, $\xi=0.1$,
    $\rho=0.03$, and $E_c=600$. The average distance between adjacent
    LSs is $0.028$. In figure (a) the curves are guide to the eye.} \label{fig4} 
\end{figure}


\subsection{Special contributions to the heat conductances}

\label{SPECIAL}

\begin{figure}[htb]
  \centering \includegraphics[height=3.9cm]{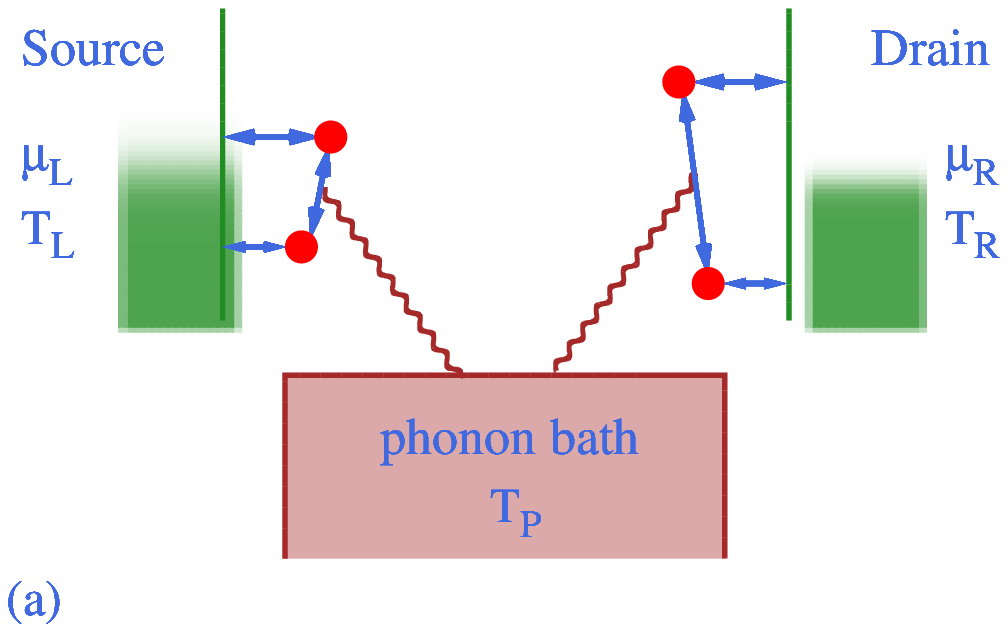}
  \centering \includegraphics[height=4.cm]{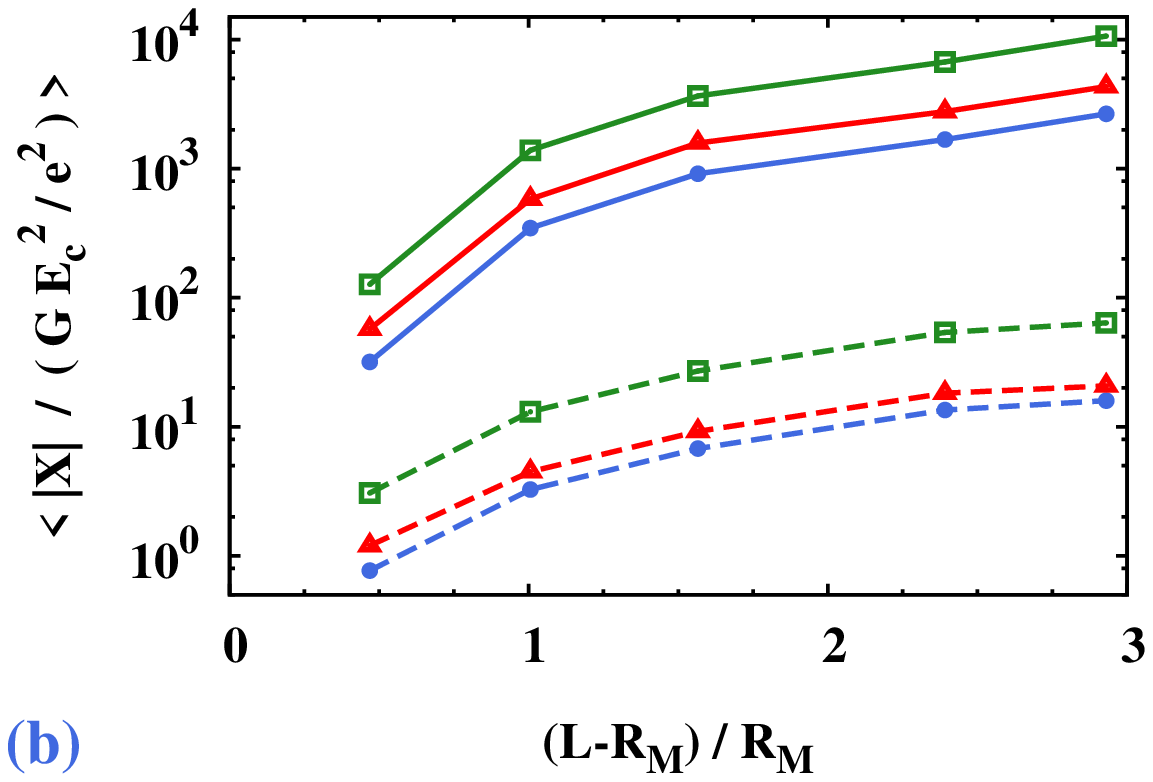}
  \centering \includegraphics[height=4.cm]{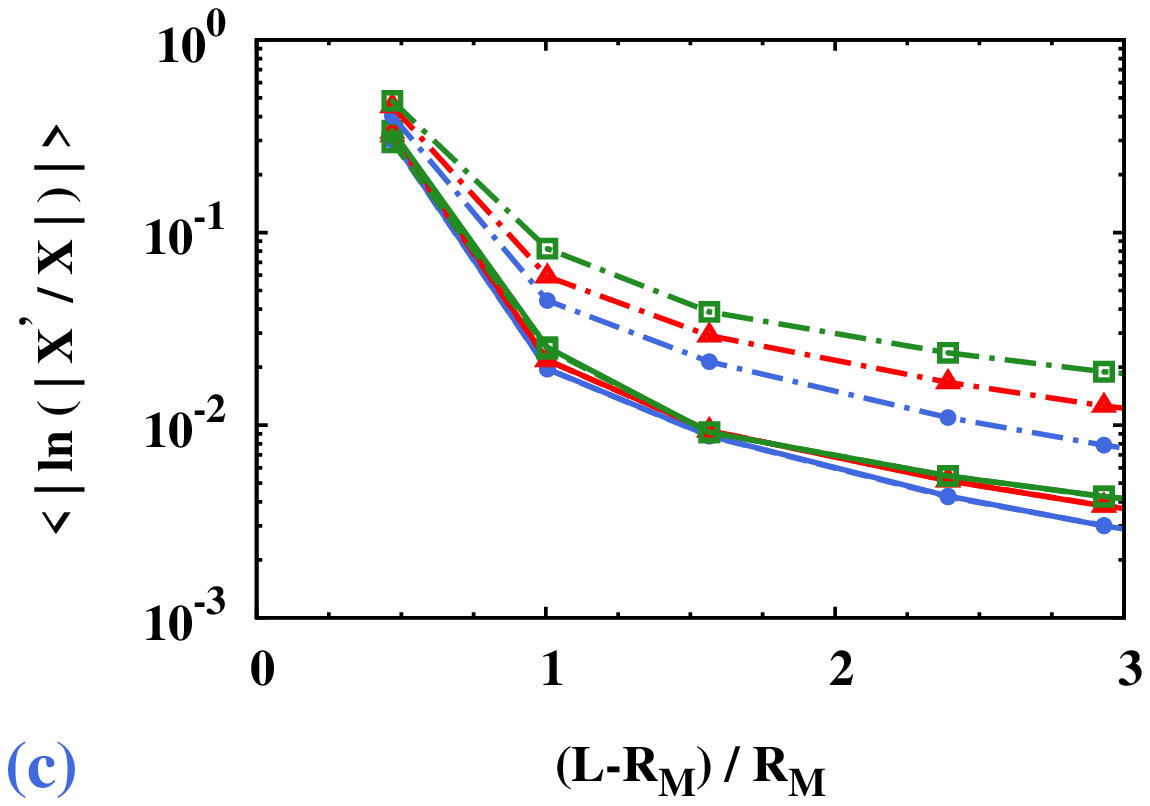}
  \caption{(Color online) (a): Schematic of the
    non-spanning hopping paths which may contribute significantly to the
    heat conductances. The wavy lines denote the phonons involved in
    the processes. (b) and (c): Distance $(L-R_M)/R_M$ dependence of
    the average of (b) $|X|/(GE_{c}^2e^{-2})$ with $X=K_e^0$
    ($\bullet$), $X=L_3$ ($\triangle$), and 
    $X=K_{pe}$ ($\square$) for $k_BT=$15 (solid curves)
    and 30 (dashed curves), (c) $|\ln(|X^{\prime}/X|)|$ for $X=K_{pe}$
    (solid curves), $X=K_e^0$ (dashed curves), and 
    $X=L_3$ (dot-dashed curves) for three 
    different temperatures with $k_BT=$15 ($\bullet$), 30
    ($\triangle$), and 60 ($\square$). The other parameters are the same as
    those in Fig.~\ref{fig3}. The results are obtained by averaging over
    $10^6$ random configurations. Note that in figure (c) the curves
    for $K_e^0$ are quite indistinguishable from those of $K_{pe}$ since
    $K_{pe}=4K_e^0$ when the non-spanning paths determine the heat
    conduction. The curves in figures (b) and (c) are guide to the eye.}\label{fig6}
\end{figure}

In deriving Eqs.~(\ref{re1}) and (\ref{re2}) we have assumed that the heat
current is carried by the percolating (spanning) paths which 
transport the charge current as well. Therefore, the heat
conductances $K_e^0$, $L_3$, and $K_{pe}$ were all proportional to the
conductance $G$. However, 
non-spanning paths
can also contribute to heat conduction. This mechanism becomes
especially important when the conductance $G$ is sufficiently small.
As the non-spanning paths do not conduct charge between the two leads,
they have no contribution to $G$, $L_1$, and $L_2$. A  
non-spanning hopping path is schematically shown in Fig.~\ref{fig6}(a). It is seen
that by hopping back and forth between a lead and the nearby LSs  having
different energies, the associated phonon energy is transferred between the lead
and the phonon bath. Therefore, whenever $T_L\ne T_P$ ($T_R\ne T_P$)
there will be heat flowing between the left (right) lead and
the phonon terminal. Even when there is no spanning hopping path
this scenario can lead to a finite heat conduction. Denoting the heat
conductances due to non-spanning paths on the left and on the right sides
by $K_L$ and $K_R$,  respectively, the corresponding heat currents are  $\dot Q_L = K_L(T_P-T_L)/T$ and
$\dot Q_R  = K_R(T_P-T_R)/T$. Hence,  when the non-spanning paths determine the
heat conduction, we find using the definitions of currents and affinities in and below Eqs.~\ref{para}
and the transport coefficients of Eq.~\ref{trans},
\bea
&& K_e^0=(K_L+K_R)/4\ ,\quad L_3 = (K_R - K_L)/2\ ,\nn\\
&&  K_{pe}=K_L+K_R=4K_e^0 \ .
\eea
In Fig.~\ref{fig6}(b) we show the numerically-computed  averages of
$K_e^0/(GE_{c}^2e^{-2})$ (curves with $\bullet$),
$|L_3|/(GE_{c}^2e^{-2})$ (curves with $\triangle$), and
$K_{pe}/(GE_{c}^2e^{-2})$ (curves with $\square$). It is seen that the
heat conductances are too large to be 
explained by the contributions from the spanning paths, for which the
upper bound is $GE_{c}^2e^{-2}$. This becomes more significant at
lower temperatures or  for longer lengths $L$, where the conductance is $G$ is
reduced but the non-spanning paths are only marginally affected. Clearly,
the Wiedemann-Franz law totally breaks down here. These results confirm that
the non-spanning paths can have much larger contributions to the heat
conductances than the spanning ones, when the conductance $G$ is
suppressed. Most of the non-spanning paths are also close
to the boundary since the sites there are much more strongly coupled to the leads and the conductance and the number of long paths
are substantially reduced.\cite{AHL} 
The boundary effect is
demonstrated in Fig.~\ref{fig6}(c). One  notes that the distance
dependencies in this case are different from those of the changes in $S$ and
$S_p$ due to the different conduction mechanisms. 
Finally, this  scenario
is not important in 1D NNH systems when the tunneling conductance  
between the left (right) lead and the LSs other than the leftmost
(rightmost) LS is small enough. It can, however, play a role in 2D NNH systems when  more than
one LSs is strongly coupled to each of the leads.

\section{Conclusions and Discussion}

\label{CD}

The study of the thermopower in the hopping regime \cite{ZO,zvyaginbook}
is augmented in this paper  in two 
ways: 1. by considering the  appropriate 
three-terminal \cite{3t,ourprb} case; and 2. by studying the 
possibly-all-important effect of the edges on the thermoelectric transport.
We emphasize that the three-terminal picture 
is dictated by the fact that hopping conductance necessitates energy-exchange 
with a (usually bosonic) thermal bath.

We studied the boundary effect on the thermoelectric properties of
finite 1D and 2D hopping systems. We find that the boundary effect
may play a crucial role for  the thermopowers. This
is first shown for a simple three-site hopping model and then for 1D
NNH systems via analytical and numerical discussions. For 1D VRH systems 
qualitative arguments and numerical
results indicate that only the LSs with a distance from the
boundaries smaller than or comparable to the Mott distance $R_M$ can affect the thermoelectric properties
considerably. As a consequence, the thermopowers  of a specific sample  of a 
very long 1D hopping system where 
the particle-hole asymmetry is negligible on average is still {\em finite and
fluctuating} due to the insufficient averaging at the boundaries. This
is confirmed by numerical calculations in 1D VRH systems where a nonzero
variance of  the thermopower persists and eventually becomes a constant at
very large system size. At the same time the average of the thermopower over 
many samples does vanish. We emphasize that the sample-dependent changes 
in the thermopower due to modifying the middle of the sample do exist, but they 
can be much smaller than the corresponding changes in the conductance.

For 2D systems we first found a situation which resembles the NNH 1D
cases: when the electronic leads are geometrically sharp and each
of them is coupled with a single LS (as with a high-resolution STM probe). In this type of systems the
boundary effect completely determines the thermoelectric
properties. However, in other types of 2D hopping systems the bulk
effect can also be important, but much less so than for the conductance. 
This is manifested by a simplified type of 2D
hopping systems which are made up of parallel 1D hopping chains where
there is no hopping between different chains. The total thermopower is
an average of the thermopower in each chain weighted by the
conductance in that chain. Changing the central part of the system will
alter the electric current in each chain  differently. Although the
thermopower in each chain does not change, the  weights of the various chains does change.
Therefore the total thermopower is modified by changing the central
part. This modification can again be much smaller than that for the conductance
(for example, when the latter is changed by many orders of magnitude  by 
changing the bulk conductance that much)
and it disappears upon ensemble averaging. In general the sample-specific 
thermoelectric properties depend mostly on the boundaries and in a limited 
fashion  on the bulk  whenever the current distribution at the
boundaries can be affected by the bulk. This includes 1D VRH systems, realistic 2D ones
(for which some numerical results are presented) and we propose also 3D ones.

For the thermal conductances sector of the linear-transport matrix [see Eq. (\ref{trans})] for the three-terminal geometry 
considered here we find new contributions. These are not due to the usual percolating paths, and 
will become important whenever the electrical conductance, due to the latter, is small enough.
This implies a very serious breakdown of the Wiedemann-Franz law. 

\section*{Acknowledgments} 
We thank Ariel Amir, Zvi Ovadyahu and Michael Pollack for illuminating discussions. OEW acknowledges the
support of the Albert Einstein Minerva Center for Theoretical Physics,
Weizmann Institute of Science. This work was supported by the BMBF
within the DIP program, BSF, by ISF, and by its Converging
Technologies Program.

\begin{appendix}

\section{Numerical scheme for solving the resistor network in
  three-terminal geometries}
\label{app:num}

Here we  present the numerical scheme for solving the resistor
network in three-terminal geometries,  in the linear-response regime. The
key quantities to be calculated are those appearing in Eq. (\ref{gg1}),    $U_i$ and  $U_{ij}$. In the linear-response regime, 
\begin{align}
&U_{ij}=\frac{|\vep_j-\vep_i|}{e}\frac{\delta T_p}{T}\ ,\nonumber\\
&U_{L,R}(\vep_i)=\frac{\delta\mu_{L,R}}{e} +
\frac{\vep_i-\mu}{e}\frac{\delta T_{L,R}}{T}\ .
\end{align}
Here $\delta\mu_i = \mu_i
-\mu$ and $\delta T_i=T_i-T$ with $i=L,R, P$. To simplify the
calculation one may choose $\mu=(\mu_L+\mu_R)/2$  
and       $T=(T_L+T_R)/2$, 
so that
$\delta\mu_L=-\delta\mu_R=\delta\mu/2$, $\delta T_L=-\delta
T_R=\delta T/2$, and $\delta T_p=\Delta T$.
According to the
sign convention in Eq.~(\ref{gg1}), $\pm U_{ij}=(\vep_j-\vep_i)\delta T_p/eT$. %
The final form of the equations to be solved is thus
\be
\sum_j A_{ij}U_j = z_i\ ,\label{z}
\ee
with
\bea
A_{ii} &=& \sum_{k\ne i}G_{ik}+G_{iL}+G_{iR} , \nn\\
A_{ij} &=& - G_{ij} \quad ({\rm for}\ i\ne j),\nn\\
z_i &=& G_{iL}\left(\frac{\delta\mu_L}{e} +
\frac{\vep_i-\mu}{e}\frac{\delta T_L}{T}\right) \nn\\
&&\mbox{} + G_{iR}\left ( \frac{\delta\mu_R}{e} +
\frac{\vep_i-\mu}{e}\frac{\delta T_R}{T} \right) \nn\\
&&\mbox{} + \sum_{k\ne i}
G_{ik}\left( \frac{\vep_i-\vep_k}{e}\frac{\delta T_p}{T} \right) .
\eea
Once the $U_i$'s are obtained by solving Eqs. (\ref{z}), the three
currents are found straightforwardly, 
\bea
I_e &=& \frac{1}{2} \sum_i (I_{i\to R} - I_{i\to L} )\ ,\nn\\
I_Q^e &=& \frac{1}{2} \sum_i (I_{i\to R} - I_{i\to L}
)\frac{\vep_i-\mu}{e}\ ,\nn\\
I_Q^{pe} &=& \sum_i (I_{i\to R} + I_{i\to L} )\frac{\vep_i-\mu}{e}\ ,
\eea
where
\bea
I_{i\to L} &=& G_{iL}\left[U_i - \left(\frac{\delta\mu_L}{e} +
\frac{\vep_i-\mu}{e}\frac{\delta T_L}{T}\right) \right] ,\nn\\
I_{i\to R} &=& G_{iR}\left[U_i - \left(\frac{\delta\mu_R}{e} +
\frac{\vep_i-\mu}{e}\frac{\delta T_R}{T}\right) \right]  .
\eea
The transport coefficients are obtained by computing the currents for three different cases:
(i) $\delta\mu\ne 0$ but $\delta T=\Delta T=0$,
(ii) $\delta T\ne 0$ but $\delta\mu=\Delta T=0$, and (iii) $\Delta
T\ne 0$ but $\delta\mu=\delta T=0$.     Using Eqs. (\ref{trans}) in case (i)
one obtains $G$, $L_1$, and
$L_2$. In case  (ii) one finds $L_1$, $K_e^0$, and $L_3$, and 
case  (iii) yields $L_2$, $L_3$, and $K_{pe}$. The
Onsager reciprocity relationships can then be verified from the numerical
computation explicitly.

\section{Comparing the conductance in the dominant hopping path with other conductances in the three-site NNH model}
\label{ap:tun}

Besides the dominant hopping path demonstrated in Fig.~\ref{fig2}(a), there are
the following transport processes: (A) elastic tunneling through
the whole system; (B) tunneling from the left lead to LS1,   hopping
from LS1 to LS2 and then tunneling into the right lead; (C)
tunneling from the left lead to LS2,   hopping to LS3 and then
tunneling into the right lead;  (D) tunneling from the left lead to LS1, 
hopping from LS1 to LS3,  and then tunneling into the right
lead. One must keep in mind the assumption that  $G_{1L}$ and $G_{3R}$ are much
larger than  all  other conductances. The conductance of (B) and (C) are   $G_B = G_{12}G_{2R}/(G_{12}+G_{2R})$
and        $G_C=G_{2L}G_{23}/(G_{2L}+G_{23})$.   
If $G_{12}\sim
G_{23}\gg G_{2R}\sim G_{2L}$ then such contributions can be
negligible. According to Eq.~(\ref{gijl}) this condition is fulfilled since 
$G_{2L}/G_{12}\simeq
\exp[-2(r_{2L}-r_{12})/\xi]=\exp(-2r_{1L}/\xi)\ll 1$ and $G_{2R}/G_{23}\simeq
\exp[-2(r_{2R}-r_{23})/\xi]=\exp(-2r_{3R}/\xi)\ll 1$. The conductance in process (D)
can be considered similarly.

It remains to consider the conductance of
process (A). The following analysis generalizes the one of
Ref.~\onlinecite{ourprb} which discusses only the two-LS assisted
tunneling [see Eq.~(\ref{2ls}) below]. The tunneling conduction consists of several
contributions. It can be assisted by one, two, or three of the three
LSs. For example,  the tunneling conductance through  LS$i=(1,2,3)$
can be written as
\be
g_i \simeq e^2 \vep_i^{-2}|\alpha_{iL}|^2|\alpha_{iR}|^2 \rho_L\rho_R\ ,
\ee
with $\rho_L$ and $\rho_R$ being the density of states in the left and
right leads,  respectively,  $|\alpha_{iL}|\simeq |\alpha_e|
\exp(-r_{iL}/\xi)$ and $|\alpha_{iR}|\simeq |\alpha_e|
\exp(-r_{iR}/\xi)$ where $|\alpha_e|$ measures the tunnel coupling
strength between the electronic states. The asymptotic behavior of
$g_i$ is thus
\be
g_i \sim e^2 \vep_i^{-2} |\alpha_{e}|^4 \rho_L\rho_R
\exp\left(-\frac{4L}{\xi}\right)\  ,
\ee
where $2L=r_{iL}+r_{iR}$ is the length of the system. Similarly one can find the
tunneling conductance through two LSs $i\ne j=(1,2,3)$ as
\be
g_{ij}\sim e^2 \vep_i^{-2} \vep_j^{-2} |\alpha_{e}|^6 \rho_L\rho_R
\exp\left(-\frac{4L}{\xi}\right)\  .\label{2ls}
\ee
Similar exponential dependence is also found for the tunneling through
three LSs. The asymptotic behavior of the total tunneling conductance
$G_{tun}$ is then
\be
G_{tun} \sim e^2 \vep_{tun}^{-2} |\alpha_{e}|^4 \rho_L\rho_R
\exp\left(-\frac{4L}{\xi}\right) \ ,
\ee
with $\vep_{tun}^{-2}=\sum_i\vep_i^{-2}+\sum_{i\ne j}\vep_i^{-2}
\vep_j^{-2} |\alpha_{e}|^2 + ...$. In comparison, the conductance of the
hopping channel is given by Eq.~(\ref{g123}). If $G_{12}\sim G_{23}$
and $|\vep_1-\mu|,|\vep_2-\mu|\gg k_BT$ then 
\bea 
G &\sim& \frac{1}{2} G_{12} \sim \frac{e^2}{k_BT} \gamma_{ep}\nn\\
&&\mbox{}\times \exp\left(
  -\frac{|\vep_1-\mu|+|\vep_2-\mu|+|\vep_1-\vep_2|}{2k_BT} - \frac{2
    r_{12}}{\xi}\right) .\nn
\eea
It is seen that the hopping conductance is limited by the exponential
factor at very low temperatures. Therefore the hopping conduction
dominates at relatively high temperatures while the tunneling is more important at
low ones. 
Ignoring the difference in the  tunnel coupling and
electron-phonon coupling, 
i.e. to a logarithmic accuracy, the crossover
temperature  between the two types of conductance, $T_x$,  is given by 
\be
k_BT_x \simeq
\frac{|\vep_1-\mu|+|\vep_2-\mu|+|\vep_1-\vep_2|}{4(2L-r_{12})} \ .
\ee

\section{A probability analysis of thermoelectric transport in
  the NNH three-site model}
\label{ap:prob}

Denoting the probability for an electron at LS 1 to be
transferred to LS 2 per unit time by $P_{1\to 2}$, and 
that of the
transfer from LS 2 to LS 3 by $P_{2\to 3}$, the entire  
probability per
unit time for an electron to be transferred from LS 1 to LS 3 by
passing LS 2 is 
\be
P_{1\to 3} = P_{1\to 2}\tilde{P}_{2\to 3} \ .
\ee
$\tilde{P}_{2\to 3}$ is the probability for the transfer from LS 2 to
LS 3 when the electron is already at LS 2. In that case there are two
possibilities: the electron can either hop to LS 1 or  to LS 3. Hence
$\tilde{P}_{2\to 3}$ is  given by the ratio $P_{2\to 3}/(P_{2\to
    1}+P_{2\to 3})$, and consequently 
%
\be
P_{1\to 3} =  \frac{P_{1\to 2}P_{2\to 3}}{P_{2\to 1}+P_{2\to 3}} \ .
\ee
The probability per unit time for the reversed process is
\be
P_{3\to 1} =  \frac{P_{3\to 2}P_{2\to 1}}{P_{2\to 1}+P_{2\to 3}} \ .
\ee
The Fermi golden-rule [see Eq. (\ref{g1})] implies that
$P_{1\to 2}=\gamma_{12}f_1(1-f_2)N_{21}$, $P_{2\to
  1}=\gamma_{12}f_2(1-f_1)(N_{21}+1)$, $P_{2\to
  3}=\gamma_{23}f_2(1-f_3)(N_{23} +1)$, and $P_{3\to
  2}=\gamma_{23}f_3(1-f_2)N_{23}$. At equilibrium $P_{1\to
  3}=P_{3\to 1}$. When the system is out of equilibrium, in the
linear-response regime, one has
\bea
I &=& P_{1\to 3} - P_{3\to 1} \nn\\
&=& \frac{P_{1\to 2}P_{2\to 3}-P_{3\to 2}P_{2\to 1}}{P_{2\to 1}+P_{2\to 3}} \nn\\
&=& \frac{G_{12}G_{23}}{G_{12}+G_{23}} [-U_1-U_{12}+U_3 +U_{23}]\nn\\
&=&\frac{G_{12}G_{23}}{G_{12}+G_{23}}\Big[\frac{\delta
  \mu}{k_BT}+\frac{\ov{\vep}_{31}}{k_BT}\frac{\delta T}{T} +
\frac{\ome_{31}}{k_BT}\frac{\Delta T}{T}\Big] .
\eea
This confirms the results obtained from the rate equation method,
i.e., Eq.~(\ref{rem}).

\end{appendix}


\begin{thebibliography}{99}
\bibitem{Com} The  thermal terminal is also  
coupled to the {\em phonons} in the  system and will determine their temperature 
if that coupling is strong enough.
\bibitem{note1} The cut-off energy $E_{c}$ is assumed  to be sufficiently higher
  than the  Mott hopping energy $E_M$ relevant for one dimension [see discussion around Eqs. (\ref{tm}), (\ref{nu}), and (\ref{e0})]. 
\bibitem{note3}The polaron effect is also not  considered in this work.
\bibitem{note6} Figure.~\ref{fig3} shows in fact   that
the thermopower is insensitive to the ``break'' created
artificially in the middle of the system.
\bibitem{note5} A small narrow peak around zero is seen in the probability
  distribution. We are not certain whether it  represents interesting
  physics or is an artifact of  the numerical procedure. This is outside of our main topic, and its study will be continued in a subsequent work.
\end{thebibliography}
\end{document}